\documentclass[aps,twocolumn,showpacs,preprintnumbers,floats]{revtex4}\def\@cite#1#2{\textsuperscript{[{#1\if@tempswa , #2\fi}]}}
\usepackage{mathrsfs}

\usepackage{longtable,lscape}
\usepackage{savesym}
\usepackage{txfonts}
\savesymbol{iint}
\savesymbol{iiint}
\savesymbol{iiiint}
\savesymbol{idotsint}
\usepackage{amssymb}
\usepackage{amsmath}
\restoresymbol{TXF}{iint}
\restoresymbol{TXF}{iiint}
\restoresymbol{TXF}{iiiint}
\restoresymbol{TXF}{idotsint}
\usepackage{indentfirst}
\usepackage{graphicx,booktabs}
\usepackage{braket}
\usepackage{multirow}
\usepackage{color}
\usepackage{subfigure}
\usepackage{float}
\usepackage{bm}
\usepackage{epsfig}
\usepackage[colorlinks, citecolor=blue,anchorcolor=red,menucolor=red, linkcolor=red,filecolor=red,urlcolor=blue,frenchlinks=red]{hyperref}
\newcommand{\vsig}{\mbox{\boldmath$\sigma$\unboldmath}}
\newcommand{\vrho}{\mbox{\boldmath$\rho$\unboldmath}}
\newcommand{\vlab}{\mbox{\boldmath$\lambda$\unboldmath}}

\begin{document}

\title{The strong decays of the low-lying $\rho$-mode $1P$-wave singly heavy baryons}
\author{Wen-Jia Wang$^{1}$, Li-Ye Xiao$^{1}$~\footnote {E-mail: lyxiao@ustb.edu.cn}, Xian-Hui Zhong$^{2,3}$~\footnote {E-mail: zhongxh@hunnu.edu.cn}}
\affiliation{ 1)Institute of Theoretical Physics, University of Science and Technology Beijing,
Beijing 100083, China}
\affiliation{ 2) Department of
Physics, Hunan Normal University, and Key Laboratory of
Low-Dimensional Quantum Structures and Quantum Control of Ministry
of Education, Changsha 410081, China }
\affiliation{ 3) Synergetic
Innovation Center for Quantum Effects and Applications (SICQEA),
Hunan Normal University, Changsha 410081, China}

\begin{abstract}
We have systematically calculated the strong decays of the low-lying $\rho$-mode $1P$-wave $\Lambda_{c(b)}$, $\Sigma_{c(b)}$, $\Xi_{c(b)}$, $\Xi^{'}_{c(b)}$, $\Omega_{c(b)}$
baryons using the chiral quark model within the $j$-$j$ coupling scheme. For the controversial states, our results indicate: (i) For the singly charmed heavy baryons, the newly observed $\Lambda_{c}(2910)^+$ is a good candidate of the $J^P=5/2^-$ state $\Lambda_c\ket{J^{P}=\frac{5}{2}^{-},2}_{\rho}$.  (ii) For the singly bottom heavy baryons, the $\Xi_{b}(6227)^-$ favors the $J^{P}=5/2^{-}$ state $\Xi_{b}\ket{J^{P}=\frac{5}{2}^{-},2}_{\rho}$. (iii) The other missing $\rho$-mode $1P$-wave excitations in $\Lambda_{b}$, $\Sigma_{c(b)}$ and $\Xi^{'}_{c(b)}$ families appear to be broad structures with $\Gamma$$\sim$(100-200) MeV, and their strong decay widths are sensitive to their masses. (iv) The $\rho$-mode $1P$-wave $\Xi_{c(b)}$ and $\Omega_{c(b)}$ baryons have a relatively narrow decay width of a few MeV or a few tens of MeV, and have a good potential to be observed in forthcoming experiments.
\end{abstract}
\maketitle

\section{Introduction}

Among the fantastic hadron zoo, the
singly heavy flavor baryon family plays an important role and is being constructed step
by step. Thanks to the large mass difference between the heavy quark ($c$ or $b$) and the light quarks, its inner structure can roughly explored within heavy quark symmetry~\cite{Grozin:1992yq,Mannel:1996rg}, which is a simpler method in hadron physics. Thus, the singly heavy baryons provide a good plot to better understand the non-perturbative nature of the quantum chromodynamics (QCD).

 Over the last few decades, great progress has been made in the detection of singly heavy baryons. Except for the ground states, more and more excitations have been found in experiments. For the baryons containing single charm quark, the excitations listed by the Particle Data Group (PDG)~\cite{ParticleDataGroup:2020ssz} include $\Lambda_c(2595)^+(\frac{1}{2}^-)$, $\Lambda_c(2625)^+(\frac{3}{2}^-)$, $\Lambda_c(2765)^+$ or $\Sigma_c(2765)^+$, $\Lambda_c(2860)^+(\frac{3}{2}^+)$, $\Lambda_c(2880)^+(\frac{5}{2}^+)$, $\Lambda_c(2940)^+(\frac{3}{2}^-)$, $\Sigma_c(2800)$, $\Xi_c(2790)(\frac{1}{2}^-)$, $\Xi_c(2815)(\frac{3}{2}^-)$, $\Xi_c(2923)$, $\Xi_c(2930)$, $\Xi_c(2970)(\frac{1}{2}^+)$, $\Xi_c(3055)$, $\Xi_c(3080)$, $\Xi_c(3123)$, $\Omega_c(3000)^0$, $\Omega_c(3050)^0$, $\Omega_c(3065)^0$, $\Omega_c(3090)$, $\Omega_c(3120)$. A few of states are not listed out since the evidence for their existence is poor and need further experimental investigation, such as $\Xi_c(2939)$ and $\Xi_c(2965)$~\cite{LHCb:2020iby}.
 Very recently, a possibly new excited $\Lambda_c$, namely $\Lambda_c(2910)$, was reported by the Belle Collaboration via investigating the $\bar{B}^0\rightarrow \Sigma_c(2455)\pi \bar{p}$ decay process~\cite{Belle:2022hnm}. The mass and width were measured to be $M=2913.8\pm5.6\pm3.8$ MeV and $\Gamma=51.8\pm20.0\pm18.8$ MeV, respectively. While, for the states containing single bottom quark, our knowledge is slightly less. The observed excitations listed by PDG are $\Lambda_b(5912)^0(\frac{1}{2}^-)$, $\Lambda_b(5920)^0(\frac{3}{2}^-)$, $\Lambda_b(6070)^0$, $\Lambda_b(6146)^0(\frac{3}{2}^+)$, $\Lambda_b(6152)^0(\frac{5}{2}^+)$, $\Sigma_b(6097)^{\pm}$, $\Xi_b(6100)^-(\frac{3}{2}^-)$, $\Xi_b(6227)^{-,0}$, $\Omega_b(6316)^- $, $\Omega_b(6330)^- $, $\Omega_b(6340)^- $ and $\Omega_b(6350)^-$~\cite{ParticleDataGroup:2020ssz}.  Recently, two new excited $\Xi_b^0$ states $\Xi_b(6327)^0$ and $\Xi_b(6333)^0$ were reported by the LHCb collaboration in the $\Lambda_b^0K^-\pi^+$ mass spectrum~\cite{LHCb:2021ssn}. Undoubtedly, these observed resonances provide good opportunities to establish the low-lying singly heavy baryon spectrum.

\begin{figure}[]
	\centering
	\centering \epsfxsize=5.0 cm \epsfbox{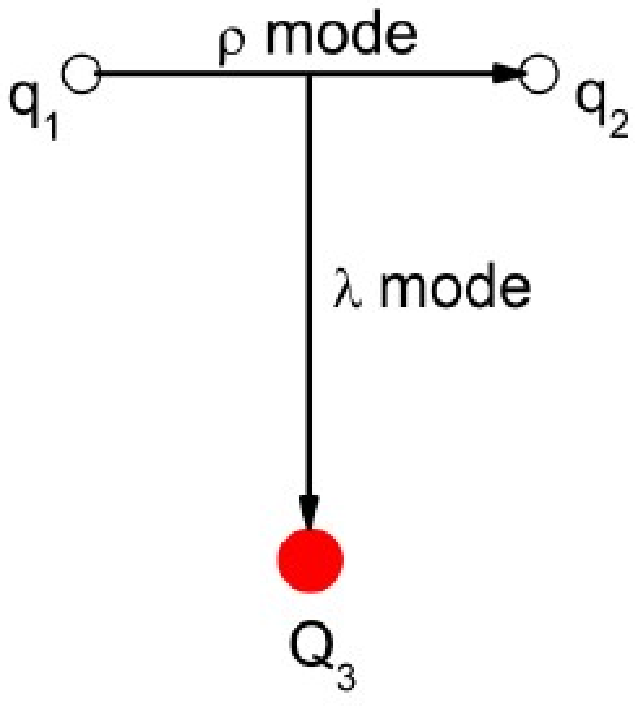}
	\caption{The $\vrho-$ and $\vlab-$ mode excitations in a singly heavy baryon. Here, $\rho$ and $\lambda$ are the Jacobi coordinates defined
		as $\vrho=\frac{1}{\sqrt{2}}(\mathbf{r}_1-\mathbf{r}_2)$ and
		$\vlab=\frac{1}{\sqrt{6}}(\mathbf{r}_1+\mathbf{r}_2-2\mathbf{r}_3)$,
		respectively. $q_1$ and $q_2$ stand for the light ($u, d, s)$ quark, and $Q_3$
		stands for a heavy charm or bottom quark. }\label{fig-1}	
\end{figure}

For a singly heavy baryon, there are two kinds of excitations:  $\rho$-mode and $\lambda$-mode excitations in theory (see Fig.~\ref{fig-1}). The $\rho$-mode excitation appears within the light diquark, while the $\lambda$-mode excitation occurs between the light diquark and the heavy quark. Using the simple harmonic oscillator model to estimate the two kinds of excitation energy~\cite{Yoshida:2015tia}, one finds that the $\rho$-mode excitation energy should be pronouncedly larger than that of the $\lambda$-mode. This indicates that the $\lambda$-mode excitations should be more easily formed than $\rho$-mode excitations. Therefore, it is natural to think that most of the low-lying observed single heavy baryons probably are good candidates of the $\lambda$-mode excitations. Meanwhile, based on the mass spectrum and strong decay analyses~\cite{Lu:2018utx,Wang:2017hej,Liu:2012sj,Guo:2019ytq,Chen:2007xf,Chen:2015kpa,Chen:2016iyi,Cheng:2017ove,Luo:2019qkm,Chen:2018orb,
Chen:2017gnu,Yang:2021lce,Roberts:2007ni,Grach:2008ij,Bijker:2020tns,Zhong:2007gp,Wang:2021bmz,Xiao:2020gjo,Wang:2020gkn,Xiao:2020oif,
Wang:2018fjm,Wang:2019uaj,Wang:2022zqv,Yoshida:2015tia,Lu:2020ivo,Kakadiya:2022zvy,Zhao:2020tpf,Suenaga:2022ajn,Azizi:2020azq,Yu:2022ymb,Yu:2021zvl}, the literatures also indicate newly observed states $\Xi_c(2923$, 2939, $2965)$~\cite{Yang:2020zjl,Wang:2020gkn,Bijker:2020tns}, $\Omega_c(3000$, $3050$, 3065, 3090, $3120)^0$~\cite{Wang:2017vnc,Chen:2017gnu,Padmanath:2017lng,Karliner:2017kfm,Wang:2017zjw}, $\Sigma_b(6097)$ and $\Xi_b(6227)$~\cite{Yang:2019cvw,Xiao:2020gjo,Cui:2019dzj,Yang:2020zrh,Aliev:2018vye,Chen:2018vuc,Yang:2018lzg,Chen:2018orb,Wang:2018fjm,Jia:2019bkr,He:2021xrh}, $\Omega_b(6316$, 6330, 6340, $6350)^- $~\cite{Wang:2020pri,Mutuk:2020rzm,Liang:2020hbo} may be explained with the $1P$-wave $\lambda$-mode excitations; $\Lambda_c(2910)$ may be the $2P$-wave $\lambda$-mode excitation~\cite{Azizi:2022dpn} and $\Lambda_c(2940)$ is a candidate of the $2P$- or $1D$-wave $\lambda$-mode excitation~\cite{Chen:2007xf,Cheng:2017ove,Guo:2019ytq,Lu:2018utx}. However, it should be mentioned that there are other explanations of those observed structures, such as $2S$-wave excitations~\cite{Agaev:2017lip,Lu:2020ivo,Agaev:2020fut,Agaev:2017jyt,Wang:2017hej,Cheng:2017ove}, $1P$-wave $\rho$-mode excitations~\cite{Bijker:2020tns,Yang:2020zrh} and unconventional interpretations~\cite{Huang:2017dwn,Liu:2018bkx,Kishore:2019fzb,Huang:2018bed,Wang:2021cku,Hu:2020zwc,Yu:2018yxl,Wang:2020vwl}.

\begin{table*}[t]
	\caption{\label{masses} The masses(MeV) of the $1P$-wave $\rho$-mode single heavy baryons obtained from the quark model.($S_{\rho}$ stands for the total spin of the two light quarks; $L$ is the total orbital angular momentum; $j$ represents the total angular momentum of $L$ and $S_{\rho}$; $J$ is the total angular momentum; $P$ is the parity; $n_{\rho}$ and $l_{\rho}$ represent the nodal quantum number and orbital angular momentum, respectively,)}
	
	\centering
	\begin{tabular}{cccccccccccc}
		\hline\hline
		state & $ n_{\rho} $ &  $l_{\rho}  $ & L & $ S_{\rho} $ &$ j $  & $ J^{P} $& CQM~\cite{Yoshida:2015tia}&RQM~\cite{Capstick:1986ter}&AQM~\cite{Roberts:2007ni}&Mass Fomula~\cite{Bijker:2020tns} &strong decay channels\\
		\hline
		$\Sigma_{c}\ket{J^{P}=\frac{1}{2}^{-},1}_{\rho}$ & 0 & 1 & 1 & 0& 1&$\frac{1}{2}^- $  & 2909&2840 & 2848&&\multirow{2}{*}{$\Sigma_{c}$$\pi$,$\Sigma^{*}_{c}$$\pi$}\\
		
		$ \Sigma_{c}\ket{J^{P}=\frac{3}{2}^{-},1}_{\rho} $ & 0 & 1 & 1 & 0& 1&$\frac{3}{2}^-$  & 2910&2865&2860& \\
		\hline
		$ \Lambda_{c}\ket{J^{P}=\frac{1}{2}^{-},0}_{\rho}$ & 0 & 1 & 1 & 1& 0&$\frac{1}{2}^- $  & 2890&2780&2816& &\multirow{5}{*}{$\Sigma_{c}$$\pi$,$\Sigma^{*}_{c}$$\pi$}\\
		
		$ \Lambda_{c}\ket{J^{P}=\frac{1}{2}^{-},1}_{\rho} $ & 0 & 1 & 1 & 1& 1&$\frac{1}{2}^- $  & 2933&2830&2816&\\
		
		$\Lambda_{c}\ket{J^{P}=\frac{3}{2}^{-},1}_{\rho} $ & 0 & 1 & 1 & 1& 1&$\frac{3}{2}^- $  & 2917&2840&2830\\
		
		$ \Lambda_{c}\ket{J^{P}=\frac{3}{2}^{-},2}_{\rho} $ & 0 & 1 & 1& 1& 2&$\frac{3}{2}^- $  & 2956&2885&2830\\
		
		$ \Lambda_{c}\ket{J^{P}=\frac{5}{2}^{-},2}_{\rho} $ & 0 & 1 & 1& 1& 2&$\frac{5}{2}^- $  & 2960&2900&2872\\
		\hline
		$\Xi^{'}_{c}\ket{J^{P}=\frac{1}{2}^{-},1}_{\rho}$ & 0 & 1 & 1 & 0& 1&$\frac{1}{2}^- $  & &&&3060&\multirow{2}{*}{$\Xi^{'}_{c}$$\pi$,$\Sigma_{c}$K,$\Xi^{*}_{c}$$\pi$,$\Sigma^{*}_{c}$K}\\
		
		$ \Xi^{'}_{c}\ket{J^{P}=\frac{3}{2}^{-},1}_{\rho} $ & 0 & 1 & 1 & 0& 1&$\frac{3}{2}^-$  & &&&3096\\
		\hline
		$ \Xi_{c}\ket{J^{P}=\frac{1}{2}^{-},0}_{\rho} $ & 0 & 1 & 1 & 1& 0&$\frac{1}{2}^- $  & &&&2951&\multirow{5}{*}{$\Xi_{c}$$\pi$,$\Xi^{'}_{c}$$\pi$,$\Sigma_{c}$K,$\Xi^{*}_{c}$$\pi$,$\Lambda_{c}$K,$\Xi_{c}(2790)$$\pi$,$\Xi_{c}(2815)$$\pi$}\\
		
		$ \Xi_{c}\ket{J^{P}=\frac{1}{2}^{-},1}_{\rho} $ & 0 & 1 & 1 & 1& 1&$\frac{1}{2}^- $  & &&&2980\\
		
		$\Xi_{c}\ket{J^{P}=\frac{3}{2}^{-},1}_{\rho} $ & 0 & 1 & 1 & 1& 1&$\frac{3}{2}^- $  & &&&2987\\
		
		$ \Xi_{c}\ket{J^{P}=\frac{3}{2}^{-},2}_{\rho} $ & 0 & 1 & 1& 1& 2&$\frac{3}{2}^- $  & &&&3016\\
		
		$ \Xi_{c}\ket{J^{P}=\frac{5}{2}^{-},2}_{\rho} $ & 0 & 1 & 1& 1& 2&$\frac{5}{2}^- $  & &&&3076\\
		\hline
		$\Omega_{c}\ket{J^{P}=\frac{1}{2}^{-},1}_{\rho}$ & 0 & 1 & 1 & 0& 1&$\frac{1}{2}^- $  & 3110&&3046& &\multirow{2}{*}{$\Xi^{'}_{c}$K}\\
		
		$ \Omega_{c}\ket{J^{P}=\frac{3}{2}^{-},1}_{\rho} $ & 0 & 1 & 1 & 0& 1&$\frac{3}{2}^-$  & 3112&&3056\\
		\hline
		$\Sigma_{b}\ket{J^{P}=\frac{1}{2}^{-},1}_{\rho}$ & 0 & 1 & 1 & 0& 1&$\frac{1}{2}^- $  & 6246&6170&6200& &\multirow{2}{*}{$\Sigma_{b}$$\pi$,$\Sigma^{*}_{b}$$\pi$}\\
		
		$ \Sigma_{b}\ket{J^{P}=\frac{3}{2}^{-},1}_{\rho} $ & 0 & 1 & 1 & 0& 1&$\frac{3}{2}^-$  & 6246&6180&6202 &\\
		\hline
		$ \Lambda_{b}\ket{J^{P}=\frac{1}{2}^{-},0}_{\rho} $ & 0 & 1 & 1 & 1& 0&$\frac{1}{2}^- $  & 6236&6100&6180&&\multirow{5}{*}{$\Sigma_{b}$$\pi$,$\Sigma^{*}_{b}$$\pi$}\\
		
		$ \Lambda_{b}\ket{J^{P}=\frac{1}{2}^{-},1}_{\rho} $ & 0 & 1 & 1 & 1& 1&$\frac{1}{2}^- $  & 6273&6165&6206&\\
		
		$\Lambda_{b}\ket{J^{P}=\frac{3}{2}^{-},1}_{\rho} $ & 0 & 1 & 1 & 1& 1&$\frac{3}{2}^- $  & 6273&6185&6211&\\
		
		$ \Lambda_{b}\ket{J^{P}=\frac{3}{2}^{-},2}_{\rho} $ & 0 & 1 & 1& 1& 2&$\frac{3}{2}^- $  & 6285&6190&6191&\\
		
		$ \Lambda_{b}\ket{J^{P}=\frac{5}{2}^{-},2}_{\rho} $ & 0 & 1 & 1& 1& 2&$\frac{5}{2}^- $  & 6289&6205&6206&\\
		\hline
		$\Xi^{'}_{b}\ket{J^{P}=\frac{1}{2}^{-},1}_{\rho}$ & 0 & 1 & 1 & 0& 1&$\frac{1}{2}^- $  & &&6305&6356&\multirow{2}{*}{$\Xi^{'}_{b}$$\pi$,$\Sigma_{b}$K,$\Xi^{*}_{b}$$\pi$,$\Sigma^{*}_{b}$K}\\
		
		$ \Xi^{'}_{b}\ket{J^{P}=\frac{3}{2}^{-},1}_{\rho} $ & 0 & 1 & 1 & 0& 1&$\frac{3}{2}^-$  &&&6308&6364\\
		\hline
		$ \Xi_{b}\ket{J^{P}=\frac{1}{2}^{-},0}_{\rho} $ & 0 & 1 & 1 & 1& 0&$\frac{1}{2}^- $  & &&&6214&\multirow{5}{*}{$\Xi_{b}$$\pi$,$\Xi^{'}_{b}$$\pi$,$\Sigma_{b}$K,$\Xi^{*}_{b}$$\pi$,$\Lambda_{b}$K,$\Sigma^{*}_{b}$K}\\
		
		$ \Xi_{b}\ket{J^{P}=\frac{1}{2}^{-},1}_{\rho} $ & 0 & 1 & 1 & 1& 1&$\frac{1}{2}^- $  & &&&6226\\
		
		$\Xi_{b}\ket{J^{P}=\frac{3}{2}^{-},1}_{\rho} $ & 0 & 1 & 1 & 1& 1&$\frac{3}{2}^- $  & &&&6222\\
		
		$ \Xi_{b}\ket{J^{P}=\frac{3}{2}^{-},2}_{\rho} $ & 0 & 1 & 1& 1& 2&$\frac{3}{2}^- $  & &&&6234\\
		
		$ \Xi_{b}\ket{J^{P}=\frac{5}{2}^{-},2}_{\rho} $ & 0 & 1 & 1& 1& 2&$\frac{5}{2}^- $  &&&&6247\\
		\hline
		$\Omega_{b}\ket{J^{P}=\frac{1}{2}^{-},1}_{\rho}$ & 0 & 1 & 1 & 0& 1&$\frac{1}{2}^- $  & 6437&&6388& &\multirow{2}{*}{$\Xi^{'}_{b}$K}\\
		
		$ \Omega_{b}\ket{J^{P}=\frac{3}{2}^{-},1}_{\rho}$ & 0 & 1 & 1 & 0& 1&$\frac{3}{2}^-$  & 6438&&6390&\\
		\hline\hline	
	\end{tabular}
\end{table*}

\begin{table*}
	\caption{\label{assignment} The parameters of the states (taken from PDG\cite{ParticleDataGroup:2020ssz}) and possible interpretations. The unit of the width is MeV.}
	\centering
	\begin{tabular}{c|c|c|c|c}
		\hline\hline
		State& Main decay channels& Total width &Possible interpretations& Our results\\
		\hline
$\Lambda_c(2910)^+$\cite{Belle:2022hnm}&$\Sigma_c(2455)^{0,++}\pi^{\pm}$&$51.8\pm20.0\pm18.8$&$2P$-wave with $J^{P}=1/2^{-}$\cite{Azizi:2022dpn} &$\Lambda_{c}$$\ket{J^{P}=5/2^{-},2}_{\rho}$\\ \hline
	\multirow{5}{*}{$\Xi_{b}(6227)^{-} $}&\multirow{5}{*}{$\Lambda_{b}$K,$\Xi_b$$\pi$}&\multirow{5}{*}{$ 18\pm6 $}&$1P$-wave $\Xi^{'}_b$ with $J^{P}=3/2^{-}$~\cite{Kakadiya:2022zvy,Chen:2018orb,He:2021xrh,Cui:2019dzj,Wang:2018fjm}  &\multirow{5}{*}{$\Xi_{b}$$\ket{J^{P}=5/2^{-},2}_{\rho}$}\\
	& & &$1P$-wave $\Xi^{'}_{b}$ with $J^{P}=5/2^{-}$~\cite{Chen:2018orb,Wang:2018fjm}&\\
		& & & $1P$-wave or $2S$-wave with  $J^{P}=3/2$~\cite{Azizi:2020azq}&\\
			& & & pentaquark molecular state with  $J^{P}=1/2^{\pm}$~\cite{Ozdem:2021vry,Wang:2020vwl,Zhu:2020lza}&\\
				& & & $\Sigma_{b}$$\overline{K}$ molecular state with  $J^{P}=1/2^{-}$~\cite{Huang:2018bed}\\
		\hline\hline

	\end{tabular}
\end{table*}

As a whole, the interpretation of the low-lying singly heavy baryons observed in recent years is still controversial. The possibility of some of those states as $\rho$-mode excitations~\cite{Bijker:2020tns,Yang:2020zrh} cannot be excluded completely. Additionally, according to the quark model predictions~\cite{Capstick:1986ter,Chen:2021eyk,Bijker:2020tns,Yoshida:2015tia} the mass of the $\rho$-mode states is about $(70-150)$ MeV higher than that of $\lambda$-mode states. We collect some theoretical predictions of the spectrum for the $1P$-wave $\rho$-mode singly heavy baryons in Table~\ref{masses}. From the table, it is seen that some observed states in experiments are in the predicted mass region of the $1P$-wave $\rho$-mode states~\cite{Yoshida:2015tia,Roberts:2007ni,Bijker:2020tns}. Besides mass spectrum, strong decay property is one of the important aspects for determining hadron's structures. Thus, to better understand the inner structures of the observed low-lying singly heavy baryons, it is crucial to study the decay properties of the $1P$-wave $\rho$-mode states. However, there are only a few discussions of the strong decays of the $1P$-wave $\rho$-mode singly heavy baryons~\cite{Yang:2020zrh,Bijker:2020tns}

In the present work, we conduct a systematical discussion of the strong decays of the low-lying $1P$-wave $\rho$-mode $\Lambda_{c(b)}$, $\Sigma_{c(b)}$, $\Xi_{c(b)}$, $\Xi'_{c(b)}$, $\Omega_{c(b)}$ baryons using the chiral quark model within the $j$-$j$ coupling scheme which includes the heavy quark symmetry. According to our theoretical calculations, we obtain that the newly observed structures $\Lambda_c(2910)^+$ and $\Xi_b(6227)^-$ may be assignments of the $1P$-wave $\rho$-mode resonances, and collect our possible explanations in Table~\ref{assignment}. We hope the predicted strong decay properties of the missing $1P$-wave $\rho$-mode singly heavy baryons be helpful for future experimental exploring.

This paper is organized as follows. In section II, we present the classification of single heavy baryons in quark model and give a brief introduction of the chiral quark model. The numerical results are presented and discussed in section III. Finally, a summary is given in section IV.

\section{ singly heavy baryons classification and Chiral quark model}

The singly heavy baryon contains a heavy quark ($c$ or $b$) and two light quarks ($u$, $d$, or $s$). The heavy quark violates the SU(4) symmetry, but the SU(3) symmetry between two light quarks is approximately kept. According to the symmetry, the singly heavy baryons belong to two different SU(3) flavor representations: the symmetry sextet $6_F$ and antisymmetric antitriplet $\overline{3}_F$.
In the singly charmed (bottom) baryons, there are two families,
$\Lambda_{c}$ and $\Xi_{c}$ ($\Lambda_{b}$ and $\Xi_{b}$) belonging to $\bar{\mathbf{3}}_F$,
while there are three families, $\Sigma_{c}$, $\Xi_{c}'$, and $\Omega_{c}$
($\Sigma_{b}$, $\Xi_{b}'$, and $\Omega_{b}$), belonging to $\mathbf{6}_F$~\cite{Wang:2017kfr}.

The spatial wave function of a singly heavy baryon is adopted the harmonic oscillator form in the constituent quark model~\cite{Zhong:2007gp}. For $q_{1}q_{2}Q$ system, it contains two light quarks $q_{1}$ and $q_{2}$ with an nearly equal constituent quark mass m, and a heavy quark $Q$ with a constituent mass $ m^{'}$. The basis states are generated by the oscillator Hamiltonian
\begin{eqnarray}
H = \frac{P^{2}_{c.m.}}{2M}+\frac{1}{2m_{\rho}}P^{2}_{\rho}+\frac{1}{2m_{\lambda}}P^{2}_{\lambda}+\frac{3}{2}K(\rho^{2}+\lambda^{2}).
\end{eqnarray}
Here, the constituent quarks are confined in an oscillator potential with the potential parameter $K$ independent on the flavor quantum number.
$\vrho$ and $\vlab$ are Jacobi coordinates as shown in Fig.~\ref{fig-1} and $\boldsymbol{R_{c.m.}}$ is the center-mass coordinate.
The momenta $ \mathbf{P}_{\rho} $, $ P_{\lambda} $ and $\mathbf{P}_{c.m.} $ are defined as
\begin{eqnarray}
\mathbf{P}_{\rho}= m_{\rho}\dot{\vrho}, ~\mathbf{P}_{\lambda}= m_{\lambda}\dot{\vlab}, ~\mathbf{P}_{c.m.}= M\dot{\mathbf{R}}_{c.m.},
\end{eqnarray}
with $M=2m+m'$, $m_\rho=m$, and $m_\lambda=\frac{3m m'}{2m+m'}$. For an oscillator, the wave function is given by
\begin{eqnarray}
\psi^{n_\sigma}_{l_{\sigma}m}(\sigma)= R_{n_{\sigma}l_{\sigma}}(\sigma)Y_{l_{\sigma}m}(\sigma)
\end{eqnarray}
where $\sigma$=$\rho$, $\lambda$. Thus, in the spatial wave functions of a singly heavy baryons, there are two oscillator parameters, i.e. the potential strengths $\alpha_{\rho}$ and $\alpha_{\lambda}$.
The parameters $\alpha_{\rho}$ and $\alpha_{\lambda}$ satisfy the following relation:
\begin{eqnarray}
\alpha^{2}_{\lambda}= \sqrt{\dfrac{3m^{'}}{2m+m^{'}}}\alpha^{2}_{\rho}.
\end{eqnarray}
The spatial wave function is a product of the $\rho$-oscillator and $\lambda$-oscillator states. With the standard notation, the principle quantum numbers of the $\rho$-mode and $\lambda$-mode oscillators are $N_{\rho}=(2n_{\rho}+l_{\rho})$ and $N_{\lambda}=(2n_{\lambda}+l_{\lambda})$, and the energy of a state is given by
\begin{eqnarray}
E= (N_{\rho} + \frac{3}{2})\omega_{\rho} + (N_{\lambda} + \frac{3}{2})\omega_{\lambda}
\end{eqnarray}
with the $\rho$-mode and $\lambda$-mode frequencies
\begin{eqnarray}
\omega_{\rho} = (3K/m_{\rho})^{1/2}, ~\omega_{\lambda} = (3K/m_{\lambda})^{1/2}.
\end{eqnarray}
Finally, the total wave function of a singly heavy baryon can be obtained, which is made up of color, spin, flavor, and spatial wave functions. Considering the color wave function is antisymmetric, the product of spin, flavor, and spatial wave functions must be symmetric. More details about the classification of the heavy baryons in the quark model can be found in Ref.~\cite{Zhong:2007gp}.

With the obtained total wave functions of the singly heavy baryons, we can further discuss their decay properties. In this work, we study the strong decay properties of the singly heavy baryons with a
chiral quark model. This model has been successfully applied
to study the strong decays of baryons and heavy-light mesons in previous works~\cite{Wang:2022zqv,Wang:2021bmz,Xiao:2020gjo,Wang:2020gkn,Xiao:2020oif,Wang:2019uaj,Wang:2018fjm}. In the chiral quark model, the effective low energy
quark-pseudoscalar-meson coupling in the SU(3) flavor basis at tree
level is described by
\begin{eqnarray}\label{STcoup}
H_m=\sum_j
\frac{1}{f_m}\bar{\psi}_j\gamma^{j}_{\mu}\gamma^{j}_{5}\psi_j\partial^{\mu}\phi_m,
\end{eqnarray}
where $\psi_j$ stands for the $j$-th quark field in a baryon. $f_m$
is the pseudoscalar meson decay constant and $\phi_m$ is the
pseudoscalar meson octet
\begin{eqnarray}
\phi_m=
\begin{pmatrix}
\frac{1}{\sqrt{2}}\pi^0+\frac{1}{\sqrt{6}}\eta & \pi^+ & K^+ \cr
\pi^- & -\frac{1}{\sqrt{2}}\pi^0+\frac{1}{\sqrt{6}}\eta & K^0 \cr
K^- & \bar{K}^0 & -\sqrt{\frac{2}{3}}
\end{pmatrix}.
\end{eqnarray}
To match the nonrelativistic harmonic oscillator spatial wave
function $^N\Psi_{LL_z}$ in this work, we adopt a nonrelativistic
form of the quark-pseudoscalar couplings with the form as follow~\cite{Li:1994cy,Li:1997gd,Zhao:2002id}
\begin{eqnarray}\label{non-relativistic-expansST}
H^{nr}_{m}=\sum_j\Big\{\frac{\omega_m}{E_f+M_f}\vsig_j\cdot
\textbf{P}_f+ \frac{\omega_m}{E_i+M_i}\vsig_j \cdot
\textbf{P}_i\\ \nonumber
-\vsig_j \cdot \textbf{q} +\frac{\omega_m}{2\mu_q}\vsig_j\cdot
\textbf{p}'_j\Big\}I_j \phi_m,
\end{eqnarray}
where the $\vsig_j$ and $\mu_q$ stand for the Pauli spin vector and
the reduced mass of the $j$-th quark in the initial and final
baryons, respectively. $\varphi_m=e^{(-)i\textbf{q}\cdot\textbf{r}_j}$ denotes (emitting)absorbing a meson. $\textbf{p}'_j=\textbf{p}_j-(m_j/M) \textbf{P}_{c.m.}$ is the
internal momentum of the $j$-th quark in the baryon rest frame.
$\omega_m$ and $\textbf{q}$ are the energy and three-vector momentum
of the meson, respectively.
$I_j$ is the isospin operator
associated with the pseudoscalar mesons.
For the emission of a light pseudoscalar meson, the partial decay
width is
\begin{equation}
\Gamma_m=\left(\frac{\delta}{f_m}\right)^2\frac{(E_f +M_f)|q|}{4\pi
	M_i}\frac{1}{2J_i+1}\sum_{J_{iz}J_{fz}}^{}|M_{J_{iz},J_{fz}}|^2,
\end{equation}
where $M_{J_{iz},J_{fz}}$ is the transition amplitude, $J_{iz}$ and
$J_{fz}$ stand for the third components of the total angular momenta
of the initial and final baryons, respectively. Accounting for the
strength of the quark-meson coupling, $\delta$ is a global parameter
which has been determined in previous study of the strong decays of
the charmed baryons and heavy-light
mesons. Here, we fix its value the
same as that in Refs.~\cite{Wang:2022zqv,Xiao:2020oif,Xiao:2020gjo,Zhong:2007gp}, i.e., $\delta=0.557$.

In the L-S coupling scheme, the states are constructed by~\cite{Roberts:2007ni}
\begin{equation}
\ket{^{2s+1}L_{J}} =\ket{[(l_{\rho}l_{\lambda})_{L}(s_{\rho}s_{Q})_{S}]_{J^P}}
\end{equation}
where $l_{\rho} $ and $l_{\lambda} $ are the quantum numbers of the orbital angular momenta for the $\rho$-mode and $\lambda$-mode (see Fig.~\ref{fig-1}) oscillators, respectively. $L$ corresponds to the quantum number of the total orbital angular momentum  $\mathbf{L}=\mathbf{l}_{\rho}+\mathbf{l}_{\lambda} $, which determines
the parity of a state by $P=(-1)^{l_{\rho}+l_{\lambda}}$. $ s_{\rho}$ and $s_{Q}$  stand for the quantum numbers of the total spin of the two light quarks and the spin of the heavy quark, respectively. $S$ is the quantum number of the total spin angular momentum $ \mathbf{S}=\mathbf{s}_{\rho}+\mathbf{s}_{Q} $. Due to the heavy quark symmetry, the physical states may be closer to the $j$-$j$ coupling scheme for the singly heavy baryons. Hence, in the present work, we study the decay properties of the singly heavy baryons in the $j$-$j$ coupling scheme.
Within the $j$-$j$ coupling scheme, the states are denoted as $\ket{J^{P},j}$ and can be expressed
as linear combination of the states within the L-S coupling
scheme by the following relationship~\cite{Roberts:2007ni}:
\begin{equation}
\begin{split}
\ket{[[(l_{\rho}l_{ \lambda })_{L}s_{\rho }]_{j}s_{Q}]_{J}}=
(-1)^{L+s_{\rho}+1/2+J}
\sqrt{2j+1} \sum_{S}\sqrt{2S+1}\\
\left(\begin{array}{ccc}
L & s_{\rho} & j\\
s_{Q} & J & S \\
\end{array} \right)
\ket{[(l_{\rho}l_{\lambda})_{L}(s_{\rho}s_{Q})_{S}]_{J^P}}.
\end{split}
\end{equation}

\begin{figure*}[]
	\centering
	\subfigure[$1P_{\rho}$-wave $\Lambda_c$ states]
	{\includegraphics[height=8.5cm,width=8.5cm]{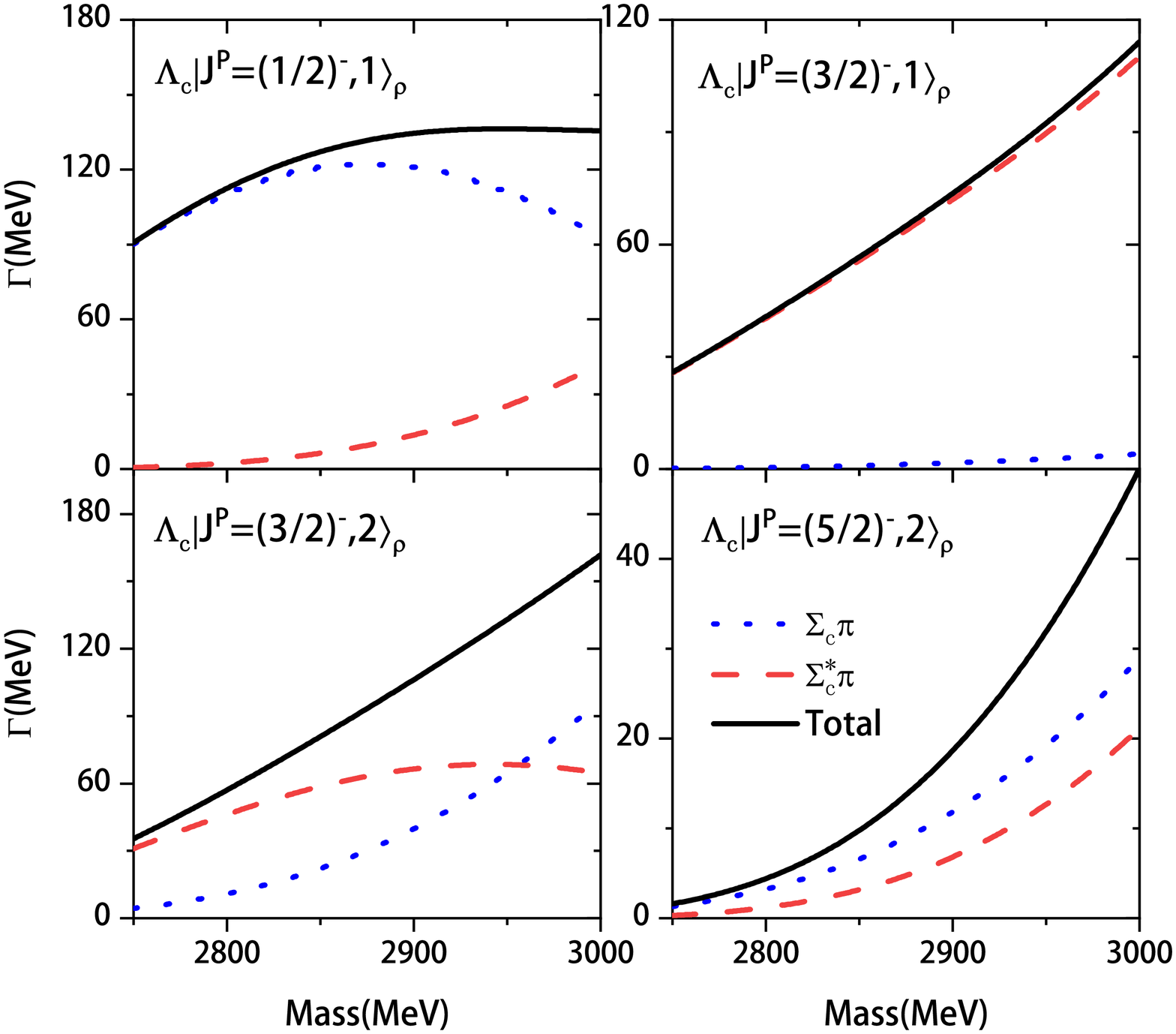}}
	\subfigure[$1P_{\rho}$-wave $\Lambda_b$ states]
	{\includegraphics[height=8.5cm,width=8.5cm]{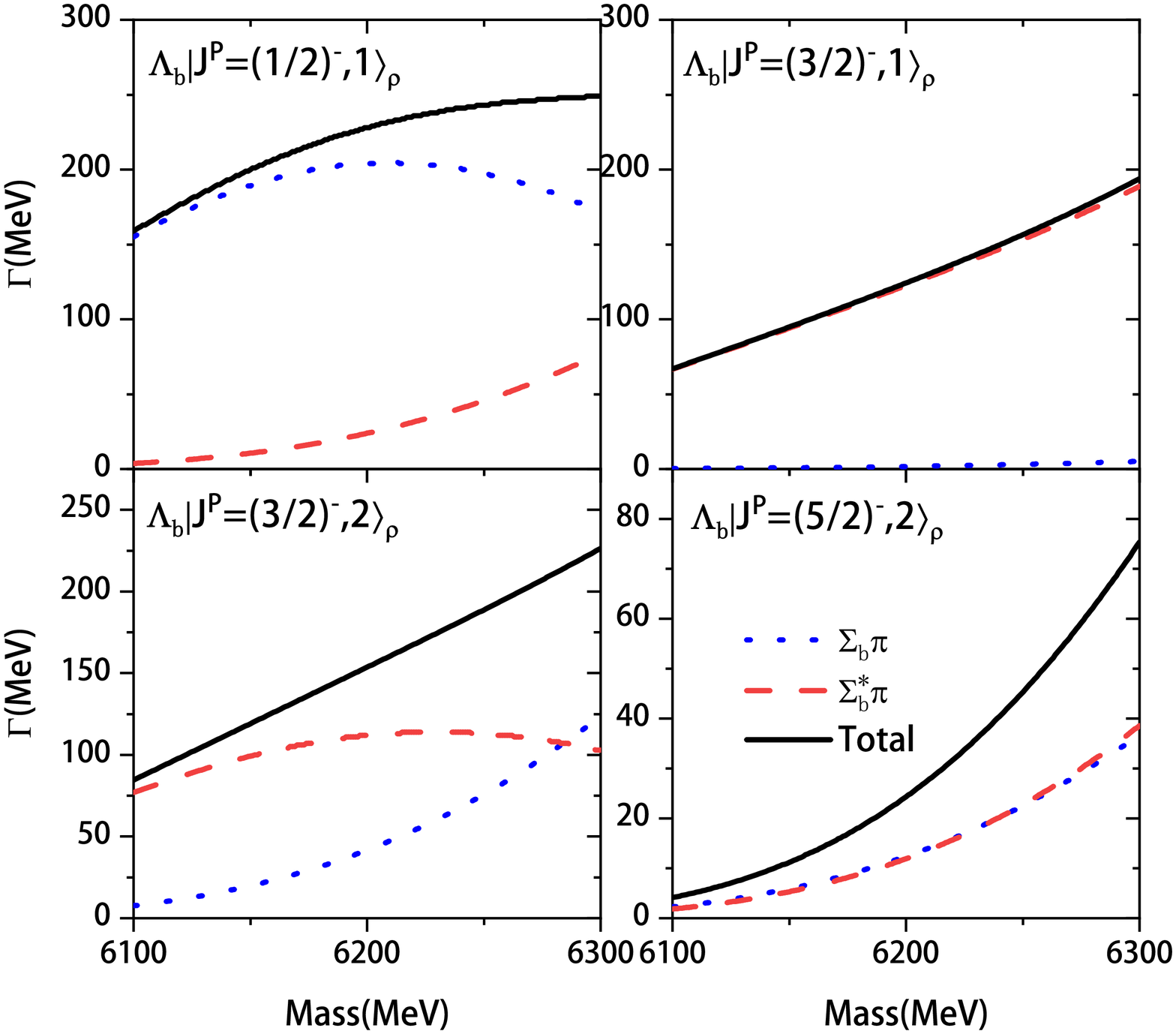}}
	\caption{Partial and total strong decay widths of the $\rho$-mode $1P$ states in the $\Lambda_{c}$ and $\Lambda_{b}$ families as functions of their masses.}\label{fig-2}
\end{figure*}

In the calculations, the standard quark model parameters are adopted. Namely, we set $ m_{u}=m_{d}=330$ MeV, $m_{s}=450 $ MeV, $m_{c}=1700 $ MeV
and $ m_{b}=5000 $ MeV for the constituent quark masses. Considering the mass differences between the u/d and s constituent quarks, the harmonic oscillator parameter $\alpha_{\rho}$ in the wave function $ ^{N}\Psi_{LL_z} $ for $uu/ud/dd$, $us/ds$ and $ss$ diquark systems should be different from each other. Thus, we take $\alpha_{\rho}$=400, 420 and 440 MeV for $uu/ud/dd$, $us/ds$ and $ss$ diquark systems, respectively~\cite{Wang:2017kfr}.
The decay constants for $\pi$ and $K$ mesons are taken as $ f_{\pi}=132 $ MeV, $ f_{K}=160 $ MeV, respectively. The masses of the well-established hadrons used in the calculations are adopted from the PDG~\cite{ParticleDataGroup:2020ssz}. With these parameters, the strong decay
properties of the well known heavy-light mesons and single heavy
baryons have been described reasonably.

\section{Calculations and Results }

Inspired by the ambiguity states observed by the collaborations, we carry out a systematic study of the strong decay behaviors of the low-lying $1P$-wave $\rho$-mode singly heavy baryons within the $j$-$j$ coupling scheme in the framework of the chiral quark model. Our theoretical results are presented as follows.

\subsection{$\Lambda_c$ and $\Lambda_b$ baryons }

In the $\Lambda_{c}$ and $\Lambda_{b}$ families, there are five $\rho$-mode $1P$ excitations
according to the quark model classification, which are
\begin{equation}
\Ket{J^{P}=\frac{1}{2}^-,0}_{\rho}=-\sqrt{\frac{1}{3}}\Ket{^2P_{\rho}\frac{1}{2}^{-}}+\sqrt{\frac{2}{3}}\Ket{^4P_{\rho}\frac{1}{2}^{-}},
\end{equation}
\begin{equation}
\Ket{J^{P}=\frac{1}{2}^-,1}_{\rho}=\sqrt{\frac{2}{3}}\Ket{^2P_{\rho}\frac{1}{2}^{-}}+\sqrt{\frac{1}{3}}\Ket{^4P_{\rho}\frac{1}{2}^{-}},
\end{equation}
\begin{equation}
\Ket{J^{P}=\frac{3}{2}^-,1}_{\rho}=-\sqrt{\frac{1}{6}}\Ket{^2P_{\rho}\frac{3}{2}^{-}}+\sqrt{\frac{5}{6}}\Ket{^4P_{\rho}\frac{3}{2}^{-}},
\end{equation}
\begin{equation}
\Ket{J^{P}=\frac{3}{2}^-,2}_{\rho}=\sqrt{\frac{5}{6}}\Ket{^2P_{\rho}\frac{3}{2}^{-}}+\sqrt{\frac{1}{6}}\Ket{^4P_{\rho}\frac{3}{2}^{-}},
\end{equation}
\begin{equation}
\Ket{J^{P}=\frac{5}{2}^-,2}_{\rho}=\Ket{^4P_{\rho}\frac{5}{2}^{-}}.
\end{equation}
 The predicted masses and possible two-body strong decay channels of these states are listed in Table~\ref{masses}.

 From the table, we can see that in the quark model the predicted masses of the $1P$-wave $\rho$-mode $\Lambda_c$ excitations are about $M\simeq(2780-3000)$ MeV, and their OZI-allowed two body strong decay channels are $\Sigma_c\pi$ and $\Sigma^*_c\pi$. However, we notice that the mass of $\Lambda_c|J^P=1/2^-,0\rangle_{\rho}$ is above the threshold of $\Sigma_c\pi$ and $\Sigma^*_c\pi$, while their strong decays are forbidden due to the orthogonality of spatial wave functions if we adopt the simple harmonic oscillator wave functions. Thus, we mainly focus on the strong decay properties of the other four states. Considering the uncertainties of the predicted masses, we plot the strong decay widths as a function of their masses in Fig.~\ref{fig-2}.

 Our results indicate that the decay properties of the four $1P$-wave $\rho$-mode $\Lambda_c$ states are sensitive to the masses. With the mass in the region of $M=(2780-3000)$ MeV, the states $\Lambda_c|J^P=\frac{1}{2}^-,1\rangle_{\rho}$, $\Lambda_c|J^P=\frac{3}{2}^-,1\rangle_{\rho}$ and $\Lambda_c|J^P=\frac{3}{2}^-,2\rangle_{\rho}$ are predicted to be moderate states with a width of $\Gamma\simeq(50-150)$ MeV. It should be pointed out that the mainly decay channel of $\Lambda_c|J^P=\frac{1}{2}^-,1\rangle_{\rho}$ is $\Sigma_c\pi$, while that of $\Lambda_c|J^P=\frac{3}{2}^-,1\rangle_{\rho}$ is $\Sigma_c^*\pi$. For the $\Lambda_c|J^P=\frac{3}{2}^-,2\rangle_{\rho}$ state, if its mass is less than $M<2958$ MeV, its dominant decay channel is $\Sigma_c^*\pi$. Otherwise its dominant decay channel is $\Sigma_c\pi$. As to the $\Lambda_c|J^P=\frac{5}{2}^-,2\rangle_{\rho}$ state, it is likely to be a narrow state with a width of $\Gamma\simeq(10-50)$ MeV when the mass varies in the region of what we considered in the present work. The mainly decay channel is $\Sigma_c\pi$, while the $\Sigma_c^*\pi$ partial decay width is considerable as well.



 Combining the natures of the newly observed state $\Lambda_c(2910)^+$ at Belle~\cite{Azizi:2022dpn}, we find that this new state may be an assignment of the narrow state $\Lambda_c|J^P=\frac{5}{2}^-,2\rangle_{\rho}$. Fixing the mass of $\Lambda_c|J^P=\frac{5}{2}^-,2\rangle_{\rho}$ at $M=2914$ MeV, the predicted total decay width
 \begin{eqnarray}
 \Gamma_{\text{Total}}\simeq22~\text{MeV},
 \end{eqnarray}
is close to the lower limit of the observed one $\Gamma_{\text{Expt.}}=51.8\pm20.0\pm18.8$ MeV. The branching fraction for the dominant decay channel $\Sigma_c\pi$ can reach up to
\begin{eqnarray}
 \frac{\Gamma[\Lambda_c|J^P=\frac{5}{2}^-,2\rangle_{\rho}\rightarrow \Sigma_c\pi]}{\Gamma_{\text{Total}}}\sim63\%.
 \end{eqnarray}
Meanwhile, the decay rate of $\Lambda_c|J^P=\frac{5}{2}^-,2\rangle_{\rho}$ into $\Sigma_c^*\pi$ is considerable, and the predicted branching fraction is
 \begin{eqnarray}
 \frac{\Gamma[\Lambda_c|J^P=\frac{5}{2}^-,2\rangle_{\rho}\rightarrow \Sigma_c^*\pi]}{\Gamma_{\text{Total}}}\sim37\%.
 \end{eqnarray}
The significant branching fraction indicates this strong decay process may be measured in future experiments and can be used to confirm the  $\Lambda_c(2910)^+$ structure as well. It should be pointed out that we cannot exclude the possibility
of $\Lambda_c(2910)^+$ as a candidate of the broader states $\Lambda_c|J^P=\frac{3}{2}^-,2\rangle_{\rho}$ and $\Lambda_c|J^P=\frac{1}{2}^-,1\rangle_{\rho}$ since there are large uncertainties in the observed width of
$\Lambda_c(2910)^+$.

In the $\Lambda_b$ family,  according to the predicted masses collected in Table~\ref{masses}, the masses of the five $\rho$-mode $1P$ $\Lambda_b$ excitations are in the region of $M\simeq(6100-6300)$ MeV.
Similarly, although the mass of the $\Lambda_b|J^P=\frac{1}{2}^-,0\rangle_{\rho}$ is above the threshold of $\Sigma_b\pi$ and $\Sigma_b^*\pi$, their strong decays are forbidden  in this work since we adopt the simple harmonic oscillator wave functions. Then, we calculate the decay properties of the other four $\rho$-mode $1P$ $\Lambda_b$ states as a function of the mass within the possible range allowed, as shown in Fig.~\ref{fig-2}. From the figure, it is found that the four $\Lambda_b$ resonances are slightly broader states compared to the corresponding states in the $\Lambda_c$ family. Meanwhile the variation curves between the partial decay width and the mass are similar to that for $\Lambda_c$ states. Among the four $\rho$-mode $1P$ $\Lambda_b$ states, $\Lambda_b|J^P=\frac{5}{2}^-,2\rangle_{\rho}$ is the narrowest state with a total width of $\Gamma_{\text{Total}}\simeq 5-74$ MeV in the range of (6100-6300) MeV. This state mainly decays into $\Sigma_b\pi$ and $\Sigma_b^*\pi$, and their partial width ratio is predicted to be
\begin{eqnarray}
 \frac{\Gamma[\Lambda_b|J^P=\frac{5}{2}^-,2\rangle_{\rho}\rightarrow \Sigma_b\pi]}{\Gamma[\Lambda_b|J^P=\frac{5}{2}^-,2\rangle_{\rho}\rightarrow \Sigma_b^*\pi]}\simeq 1.
 \end{eqnarray}
The $\Lambda_b|J^P=\frac{5}{2}^-,2\rangle_{\rho}$ state may be observed in the $\Lambda_b\pi\pi$ final state via the decay chains $\Lambda_b|J^P=\frac{5}{2}^-,2\rangle_{\rho}\rightarrow \Sigma_b\pi/\Sigma_b^*\pi\rightarrow\Lambda_b\pi\pi$.
The $\Lambda_b|J^P=\frac{1}{2}^-,1\rangle_{\rho}$ and $\Lambda_b|J^P=\frac{3}{2}^-,1\rangle_{\rho}$
dominantly decay into $\Sigma_b\pi$ and $\Sigma_b^*\pi$, respectively, they have a comparable a width of
$\Gamma\simeq 120-200$ MeV. The $\Lambda_b|J^P=\frac{3}{2}^-,2\rangle_{\rho}$ mainly decay into
$\Sigma_b^*\pi$ and $\Sigma_b\pi$ channels, and their partial width ratio is predicted to be
\begin{eqnarray}
 \frac{\Gamma[\Lambda_b|J^P=\frac{3}{2}^-,2\rangle_{\rho}\rightarrow \Sigma_b\pi]}{\Gamma[\Lambda_b|J^P=\frac{3}{2}^-,2\rangle_{\rho}\to \Sigma_b^*\pi]}\simeq 0.4-1.2.
 \end{eqnarray}

\subsection{$\Sigma_c$ and $\Sigma_b$ baryons }

\begin{table}[]
	\caption{\label{Sigma}Partial decay widths(MeV) and branching fractions for the $\rho$-mode $1P$-wave states in the $\Sigma_{c}$ and $\Sigma_{b}$ families. The numbers in parentheses stand for the corresponding masses(MeV).}
	\centering
	\begin{tabular}{l|cc|cc}
		\hline \hline
		\multirow{2}{*}{ }&\multicolumn{2}{c}{$\Sigma_{c}\ket{J^{P}=\frac{1}{2}^{-},1}_{\rho}$(2909)}
		~~~& \multicolumn{2}{c}{$\Sigma_{c}\ket{J^{P}=\frac{3}{2}^{-},1}_{\rho}$(2910)}\\
		\cline{2-5}
		&$\Gamma_i$ & $B_i(\%)$&$\Gamma_i$ & $B_i(\%)$\\
		\hline
		$\Sigma_{c}$$\pi$& 100&66&55&30\\
		$\Sigma^{*}_{c}$$\pi$&51&34 &127&70\\
		\hline
		Total&\multicolumn{2}{c}{151}&\multicolumn{2}{c}{182}\\
		\hline\hline
		\multirow{2}{*}{ }&\multicolumn{2}{c}{$\Sigma_{b}\ket{J^{P}=\frac{1}{2}^{-},1}_{\rho}$(6246)}
		~~~& \multicolumn{2}{c}{$\Sigma_{b}\ket{J^{P}=\frac{3}{2}^{-},1}_{\rho}$(6246)}\\
		\cline{2-5}
		&$\Gamma_i$ & $B_i(\%)$&$\Gamma_i$ & $B_i(\%)$\\
		\hline
		$\Sigma_{b}$$\pi$& 99&53&54&27\\
		$\Sigma^{*}_{b}$$\pi$&87&47 &145&73\\
		\hline
		Total&\multicolumn{2}{c}{145}&\multicolumn{2}{c}{199}\\
		\hline\hline		
	\end{tabular}
\end{table}

\begin{figure}[]
	\centering
	\subfigure[$1P_{\rho}$-wave $\Sigma_c$ states]
	{\includegraphics[height=7cm,width=4.1cm]{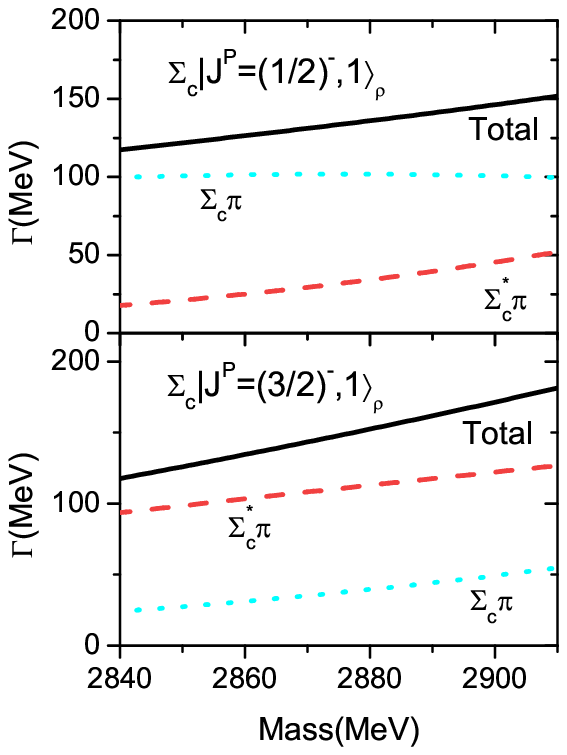}}
	\subfigure[$1P_{\rho}$-wave $\Sigma_b$ states]
	{\includegraphics[height=7cm,width=4.0cm]{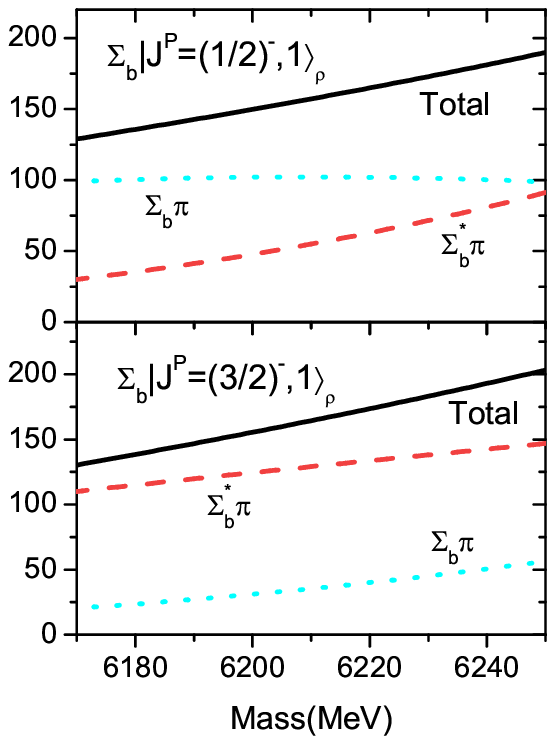}}
	\caption{Partial and total strong decay widths of the $\rho$-mode $1P$ states in the $\Sigma_{c}$ and $\Sigma_{b}$ families as functions of their masses.}\label{fig-3}
\end{figure}

In the $\Sigma_{c}$ and $\Sigma_{b}$ families, there are two $\rho$-mode $1P$-wave excitations:
\begin{equation}
\Ket{J^{P}=\frac{1}{2}^-,1}_{\rho}=\Ket{^2P_{\rho}\frac{1}{2}^{-}},
\end{equation}
\begin{equation}
\Ket{J^{P}=\frac{3}{2}^-,1}_{\rho}=\Ket{^2P_{\rho}\frac{3}{2}^{-}}.
\end{equation}
The masses of the $\rho$-mode $1P$-wave $\Sigma_{c}$ and $\Sigma_{b}$ excitations are predicted to be $M\simeq(2840-2910)$ MeV and $M\simeq(6170-6250)$  MeV, respectively. Adopting the predicted masses from Ref.~\cite{Yoshida:2015tia}, we collect their decay properties in Table~\ref{Sigma}. It is shown that all of the $\rho$-mode $1P$-wave excitations may be broad states with a total decay width of around $\Gamma\sim(150-200)$ MeV.

In the $\Sigma_c$ family, the dominant decay channel of $\Sigma_{c}\ket{J^{P}=\frac{1}{2}^{-},1}_{\rho}$ is $\Sigma_c\pi$, while that of $\Sigma_{c}\ket{J^{P}=\frac{3}{2}^{-},1}_{\rho}$ is $\Sigma_c^*\pi$. The predicted branching fractions are
\begin{eqnarray}
\frac{\Sigma_{c}\ket{J^{P}=\frac{1}{2}^{-},1}_{\rho}\rightarrow \Sigma_c\pi}{\Gamma_{\text{Total}}}\sim66\%,
\end{eqnarray}
\begin{eqnarray}
\frac{\Sigma_{c}\ket{J^{P}=\frac{3}{2}^{-},1}_{\rho}\rightarrow \Sigma_c^*\pi}{\Gamma_{\text{Total}}}\sim70\%,
\end{eqnarray}
which can be used to distinguish $\Sigma_{c}\ket{J^{P}=\frac{1}{2}^{-},1}_{\rho}$ from $\Sigma_{c}\ket{J^{P}=\frac{3}{2}^{-},1}_{\rho}$ in future experiments.

In the $\Sigma_b$ family,
the $J^P=1/2^-$ state $\Sigma_{b}\ket{J^{P}=\frac{1}{2}^{-},1}_{\rho}$ mainly decays into $\Sigma_b\pi$ and $\Sigma_b^*\pi$. Their branching fractions are
\begin{eqnarray}
\frac{\Sigma_{b}\ket{J^{P}=\frac{1}{2}^{-},1}_{\rho}\rightarrow \Sigma_b\pi}{\Gamma_{\text{Total}}}\sim53\%,
\end{eqnarray}
\begin{eqnarray}
\frac{\Sigma_{b}\ket{J^{P}=\frac{1}{2}^{-},1}_{\rho}\rightarrow \Sigma_b^*\pi}{\Gamma_{\text{Total}}}\sim47\%.
\end{eqnarray}
The $J^P=3/2^-$ state $\Sigma_{b}\ket{J^{P}=\frac{3}{2}^{-},1}_{\rho}$ is governed by $\Sigma_b^*\pi$, and branching fraction is predicted to be
\begin{eqnarray}
\frac{\Sigma_{b}\ket{J^{P}=\frac{3}{2}^{-},1}_{\rho}\rightarrow \Sigma_b^*\pi}{\Gamma_{\text{Total}}}\sim73\%.
\end{eqnarray}

%
%
Considering the uncertainty of the mass predictions of the $\rho$-mode $1P$-wave $\Sigma_{c}$ and $\Sigma_{b}$ excitations, we plot the strong decay widths as a function of the mass in Fig.~\ref{fig-3}. The sensitivities of the decay properties of those states to their masses can be clearly seen from the figure.

\begin{table*}[]
	\caption{\label{Xi}The strong decay properties of the $\rho$-mode $1P$-wave states in the $\Xi_{c}$ and $\Xi_{b}$ families.$\Gamma_{\text{Total}}$ stands for the total decay width. The unit of the width and mass is MeV. The masses for the unestablished $\rho$-mode $1P$ $\Xi_{c}$ and $\Xi_{b}$ states are taken from the predictions in Ref.~\cite{Bijker:2020tns}.}
	\centering
	\begin{tabular}{ccccccccc}	
		\hline\hline
		\multirow{2}{*}{Decay width}&$\underline{~~\Xi_{c}\ket{J^{P}=\frac{1}{2}^{-},0}_{\rho}~~}$ &$\underline{~~\Xi_{c}\ket{J^{P}=\frac{1}{2}^{-},1}_{\rho}~~}$
&$\underline{~~\Xi_{c}\ket{J^{P}=\frac{3}{2}^{-},1}_{\rho}~~}$&$\underline{~~\Xi_{c}\ket{J^{P}=\frac{3}{2}^{-},2}_{\rho}~~}$
&$\underline{~~\Xi_{c}\ket{J^{P}=\frac{5}{2}^{-},2}_{\rho}~~}$\\
		&$M$=2951&$M$=2980&$M$=2987&$M$=3016&$M$=3076\\
		\hline
		$\Gamma[\Xi_c\pi]$&48.6&0.0&10.7&10.8&37.1\\
		$\Gamma[\Lambda_{c}K]$&64.2&0.0&5.1&5.5&20.7\\
		$\Gamma[\Xi^{'}_c\pi]$&0.0&33.5&0.3&10.2&5.5\\
		$\Gamma[\Sigma_{c}K]$&0.0&102.9&0.0&1.9&2.9\\
		$\Gamma[\Xi^{*}_c\pi]$&0.0&1.9&16.1&17.9&3.5\\
		$\Gamma[\Sigma^{*}_cK]$&-&-&-&11.5&0.6\\
        $\Gamma[\Xi_c(2790)\pi]$&0.0&0.0&0.0&0.0&0.1\\
		$\Gamma[\Xi_c(2815)\pi]$&0.0&0.0&0.0&0.0&0.1\\
		$\Gamma_{\text{Total}}$&112.8&138.3&32.2&57.8&70.5\\
		\hline \hline
		\multirow{2}{*}{Decay width}&$\underline{~~\Xi_{b}\ket{J^{P}=\frac{1}{2}^{-},0}_{\rho}~~}$
&$\underline{~~\Xi_{b}\ket{J^{P}=\frac{1}{2}^{-},1}_{\rho}~~}$
&$\underline{~~\Xi_{b}\ket{J^{P}=\frac{3}{2}^{-},1}_{\rho}~~}$&$\underline{~~\Xi_{b}\ket{J^{P}=\frac{3}{2}^{-},2}_{\rho}~~}$
&$\underline{~~\Xi_{b}\ket{J^{P}=\frac{5}{2}^{-},2}_{\rho}~~}$\\
		&$M=6214$&$M=6226$&$M=6222$&$M=6234$&$M=6247$\\
		\hline
		$\Gamma[\Xi_b\pi]$&50.6&0.0&5.5&5.1&13.1\\
		$\Gamma[\Lambda_{b}K]$&76.0&0.0&1.5&1.5&4.4\\
		$\Gamma[\Xi^{'}_b\pi]$&0.0&25.0&0.1&1.5&0.6\\
		$\Gamma[\Sigma_{b}K]$&0.0&0.0&0.0&0.0&0.0\\
		$\Gamma[\Xi^{*}_b\pi]$&0.0&0.8&11.2&13.7&0.6\\
		$\Gamma[\Sigma^{*}_{b}K]$&0.0&0.0&0.0&0.0&0.0\\
		$\Gamma_{\text{Total}}$&126.6&25.8&18.3&21.9&18.7\\
		\hline\hline		
	\end{tabular}	
\end{table*}

\begin{figure*}[]
	\centering
	\subfigure[$1P_{\rho}$-wave $\Xi_c$ states]
	{\includegraphics[height=11cm,width=8cm]{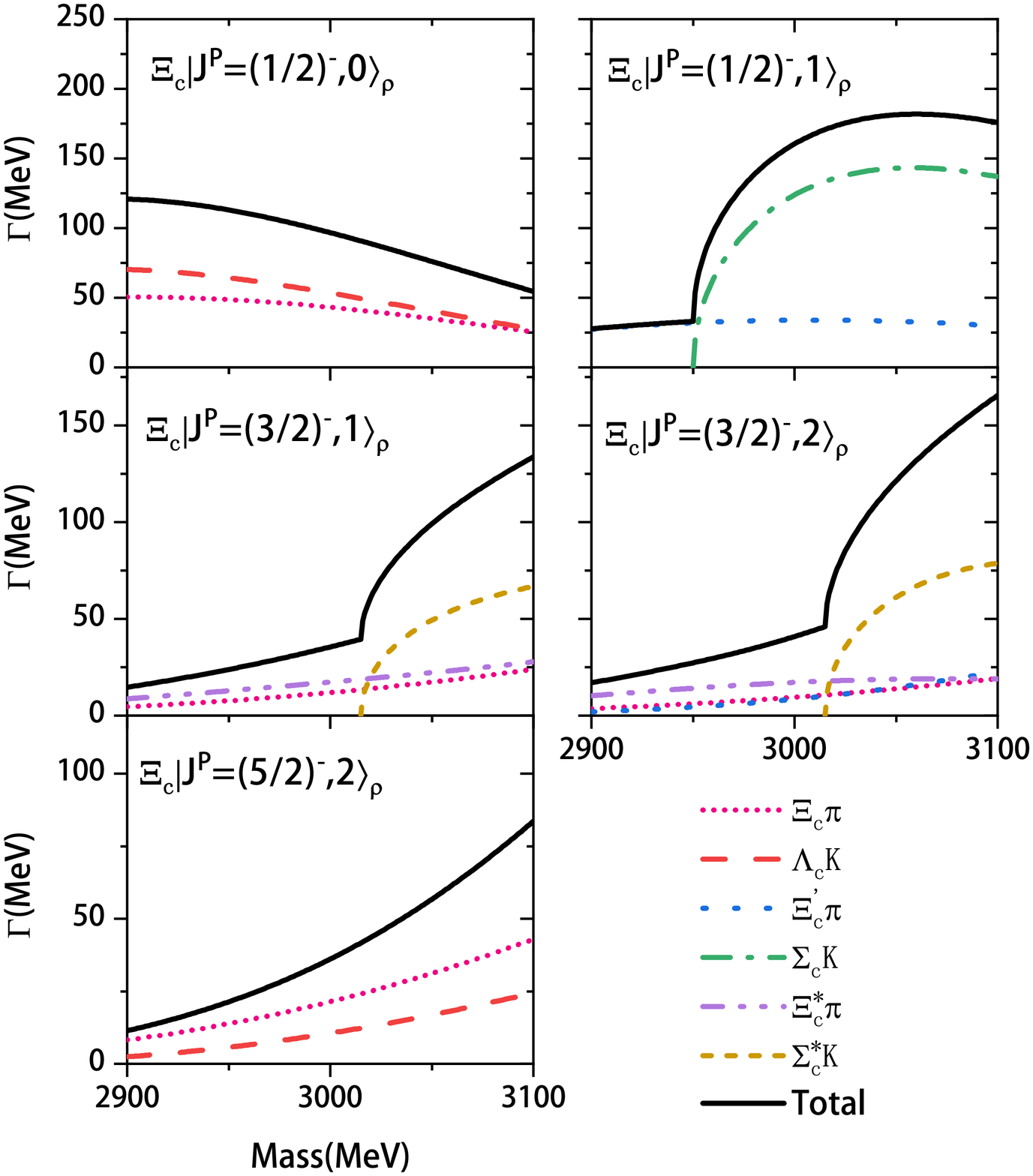}}
	\subfigure[$1P_{\rho}$-wave $\Xi_b$ states]
	{\includegraphics[height=11cm,width=8cm]{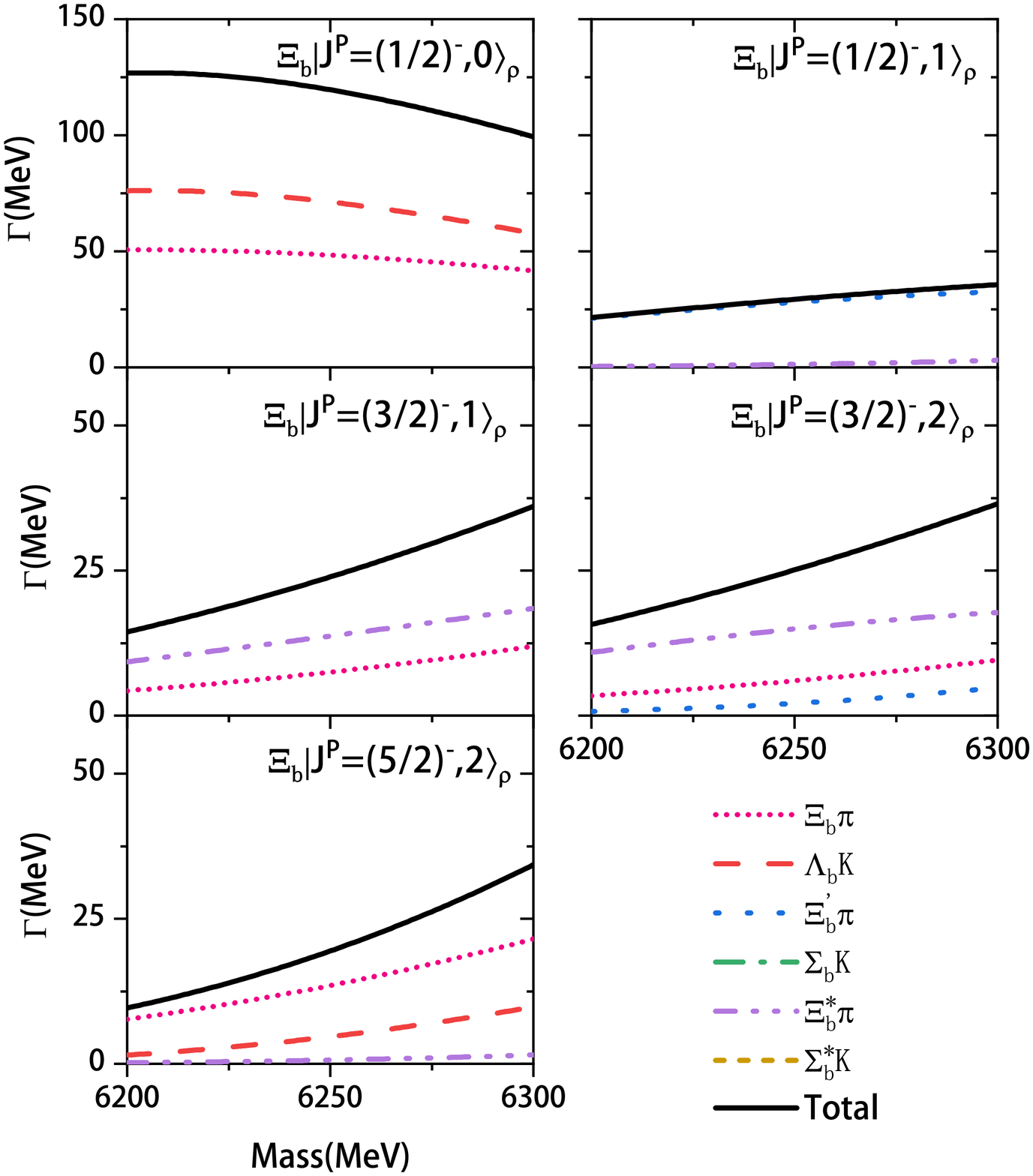}}
	\caption{Partial and total strong decay width of the $\rho$-mode $1P$ states in the $\Xi_{c}$ and $\Xi_{b}$ families as functions of their masses. Some decay channels are not shown in the figure for their small partial decay widths.}\label{fig-4}
\end{figure*}

\subsection{$\Xi_c$ and $\Xi_b$ baryons }

In the $\Xi_c$ and $\Xi_{b}$ families, there are each five $\rho$-mode $1P$-wave excitations:
$\ket{J^{P}=\frac{1}{2}^{-},0}_{\rho}$, $\ket{J^{P}=\frac{1}{2}^{-},1}_{\rho}$, $\ket{J^{P}=\frac{3}{2}^{-},1}_{\rho}$, $\ket{J^{P}=\frac{3}{2}^{-},2}_{\rho}$ and $\ket{J^{P}=\frac{5}{2}^{-},2}_{\rho}$. For their masses, there are a few discussions in theoretical references and we have collected in Table~\ref{masses}. From the table, the typical masses of the $\rho$-mode $1P$-wave $\Xi_c$ and $\Xi_{b}$ states are about $M\sim$2.9 and $M\sim$6.2 GeV, respectively. Adopting the predicted masses based on Mass Fomula~\cite{Bijker:2020tns}, we calculate their two-body strong decay properties and list in Table~\ref{Xi}.

In the $\Xi_c$ family, the $J^P=1/2^-$ state $\Xi_{c}\ket{J^{P}=\frac{1}{2}^{-},0}_{\rho}$ has a moderate width of $\Gamma\simeq113$ MeV, and dominantly decays into $\Xi_c\pi$ and $\Lambda_cK$ with predicted branching fractions
\begin{eqnarray}
\frac{\Gamma[\Xi_{c}\ket{J^{P}=\frac{1}{2}^{-},0}_{\rho}\rightarrow \Xi_c\pi]}{\Gamma_{\text{Total}}}\sim43\%,
\end{eqnarray}
\begin{eqnarray}
\frac{\Gamma[\Xi_{c}\ket{J^{P}=\frac{1}{2}^{-},0}_{\rho}\rightarrow \Lambda_cK]}{\Gamma_{\text{Total}}}\sim57\%.
\end{eqnarray}
The $\Xi_c\pi$ and $\Lambda_cK$ channels can be used to search for the missing $\Xi_{c}\ket{J^{P}=\frac{1}{2}^{-},0}_{\rho}$ state.

For the other $J^P=1/2^-$ state $\Xi_{c}\ket{J^{P}=\frac{1}{2}^{-},1}_{\rho}$, its width is predicted to be around $\Gamma\simeq138$ MeV. This staes
mainly decays into $\Xi_c'\pi$ and $\Sigma_cK$. Their branching fractions are
\begin{eqnarray}
\frac{\Gamma[\Xi_{c}\ket{J^{P}=\frac{1}{2}^{-},1}_{\rho}\rightarrow \Xi_c'\pi]}{\Gamma_{\text{Total}}}\sim24\%,
\end{eqnarray}
\begin{eqnarray}
\frac{\Gamma[\Xi_{c}\ket{J^{P}=\frac{1}{2}^{-},1}_{\rho}\rightarrow \Sigma_cK]}{\Gamma_{\text{Total}}}\sim74\%.
\end{eqnarray}

The $J^P=3/2^-$ state $\Xi_{c}\ket{J^{P}=\frac{3}{2}^{-},1}_{\rho}$ has a narrow width of $\Gamma\sim32$ MeV, and dominantly decays into the $\Xi_c\pi$, $\Lambda_cK$ and $\Xi_c^*\pi$ channels. The branching fractions are predicted to be
\begin{eqnarray}
\frac{\Gamma[\Xi_{c}\ket{J^{P}=\frac{3}{2}^{-},1}_{\rho}\rightarrow \Xi_c\pi]}{\Gamma_{\text{Total}}}\sim33\%,\\
\frac{\Gamma[\Xi_{c}\ket{J^{P}=\frac{3}{2}^{-},1}_{\rho}\rightarrow \Lambda_cK]}{\Gamma_{\text{Total}}}\sim16\%,\\
\frac{\Gamma[\Xi_{c}\ket{J^{P}=\frac{3}{2}^{-},1}_{\rho}\rightarrow \Xi_c^*\pi]}{\Gamma_{\text{Total}}}\sim50\%.
\end{eqnarray}
This state have a large potential to be observed in the $\Xi_c\pi$, $\Lambda_cK$ and $\Xi_c^*\pi$ channels.

The other $J^P=3/2^-$ state $\Xi_{c}\ket{J^{P}=\frac{3}{2}^{-},2}_{\rho}$ is slightly broader than $\Xi_{c}\ket{J^{P}=\frac{3}{2}^{-},1}_{\rho}$, and has a width of $\Gamma\sim58$ MeV. Except the $\Xi_c\pi$, $\Lambda_cK$ and $\Xi_c^*\pi$ channels, this state also has large decay rates into  $\Xi_c'\pi$ and $\Sigma_c^*K$. The branching fractions for the $\Xi_c^*\pi$ and $\Sigma_c^*K$ channels can reach up to
 \begin{eqnarray}
\frac{\Gamma[\Xi_{c}\ket{J^{P}=\frac{3}{2}^{-},2}_{\rho}\rightarrow \Xi_c^*\pi]}{\Gamma_{\text{Total}}}\sim31\%,\\
\frac{\Gamma[\Xi_{c}\ket{J^{P}=\frac{3}{2}^{-},2}_{\rho}\rightarrow \Sigma_c^*K]}{\Gamma_{\text{Total}}}\sim20\%.
\end{eqnarray}
The $\Xi_{c}\ket{J^{P}=\frac{3}{2}^{-},2}_{\rho}$ may be observed in the $\Xi_c\pi\pi/\Lambda_c\pi K$ final states via the decay chains $\Xi_{c}\ket{J^{P}=\frac{3}{2}^{-},2}_{\rho}\rightarrow \Xi_c^*\pi/\Sigma_c^*K\rightarrow \Xi_c\pi\pi/\Lambda_c\pi K$.

The decay of $\Xi_{c}\ket{J^{P}=\frac{5}{2}^{-},2}_{\rho}$ is governed by $\Xi_c\pi$ and $\Lambda_cK$ with branching fractions
 \begin{eqnarray}
\frac{\Gamma[\Xi_{c}\ket{J^{P}=\frac{5}{2}^{-},2}_{\rho}\rightarrow \Xi_c\pi]}{\Gamma_{\text{Total}}}\sim53\%,\\
\frac{\Gamma[\Xi_{c}\ket{J^{P}=\frac{5}{2}^{-},2}_{\rho}\rightarrow \Lambda_cK]}{\Gamma_{\text{Total}}}\sim29\%.
\end{eqnarray}

In the $\Xi_b$ family, the $\Xi_{b}\ket{J^{P}=\frac{1}{2}^{-},0}_{\rho}$ has a moderate width of $\Gamma\simeq127$ MeV, and mainly decays via $\Xi_b\pi$ and $\Lambda_bK$. Their predicted branching fractions are
\begin{eqnarray}
\frac{\Gamma[\Xi_{b}\ket{J^{P}=\frac{1}{2}^{-},0}_{\rho}\rightarrow \Xi_b\pi]}{\Gamma_{\text{Total}}}\sim40\%,\\
\frac{\Gamma[\Xi_{b}\ket{J^{P}=\frac{1}{2}^{-},0}_{\rho}\rightarrow \Lambda_bK]}{\Gamma_{\text{Total}}}\sim60\%.
\end{eqnarray}

The other four $\rho$-mode $1P$ states $\Xi_{b}\ket{J^{P}=\frac{1}{2}^{-},1}_{\rho}$, $\Xi_{b}\ket{J^{P}=\frac{3}{2}^{-},1}_{\rho}$, $\Xi_{b}\ket{J^{P}=\frac{3}{2}^{-},2}_{\rho}$ and $\Xi_{b}\ket{J^{P}=\frac{5}{2}^{-},2}_{\rho}$ have a comparable width of $\Gamma\sim20$ MeV. While, we notice that the main decay channels have big difference among those four states. The $\Xi_{b}\ket{J^{P}=\frac{1}{2}^{-},1}_{\rho}$ is mostly saturated by the decay channel $\Xi_b'\pi$ and the branching fraction for the $\Xi_b'\pi$ channel can reach up to
\begin{eqnarray}
\frac{\Gamma[\Xi_{b}\ket{J^{P}=\frac{1}{2}^{-},1}_{\rho}\rightarrow \Xi_b'\pi]}{\Gamma_{\text{Total}}}\sim97\%.
\end{eqnarray}

The $J^P=3/2^-$ states $\Xi_{b}\ket{J^{P}=\frac{3}{2}^{-},1}_{\rho}$ and $\Xi_{b}\ket{J^{P}=\frac{3}{2}^{-},2}_{\rho}$ dominantly decay into the $\Xi_b\pi$ and $\Xi_b^*\pi$ channels. The branching fractions are predicted to be
\begin{eqnarray}
\frac{\Gamma[\Xi_{b}\ket{J^{P}=\frac{3}{2}^{-},1(2)}_{\rho}\rightarrow \Xi_b\pi]}{\Gamma_{\text{Total}}}\sim30\%(23\%),\\
\frac{\Gamma[\Xi_{b}\ket{J^{P}=\frac{3}{2}^{-},1(2)}_{\rho}\rightarrow \Xi_b^*\pi]}{\Gamma_{\text{Total}}}\sim61\%(63\%).
\end{eqnarray}

For the $\Xi_{b}\ket{J^{P}=\frac{5}{2}^{-},2}_{\rho}$ state, the dominant decay modes is $\Xi_b\pi$ with a branching fraction of
\begin{eqnarray}
\frac{\Gamma[\Xi_{b}\ket{J^{P}=\frac{5}{2}^{-},2}_{\rho}\rightarrow \Xi_b\pi]}{\Gamma_{\text{Total}}}\sim70\%.
\end{eqnarray}
Meanwhile, the $\Xi_{b}\ket{J^{P}=\frac{5}{2}^{-},2}_{\rho}$ state has a sizeable decay rate into $\Lambda_bK$, and the branching fraction is about
\begin{eqnarray}
\frac{\Gamma[\Xi_{b}\ket{J^{P}=\frac{5}{2}^{-},2}_{\rho}\rightarrow \Lambda_bK]}{\Gamma_{\text{Total}}}\sim24\%.
\end{eqnarray}

\begin{table}[]
	\caption{\label{Xib}The partial decay widths(MeV) of $\Xi_b(6227)^-$ assigned as $\rho$-mode $1P$ $\Xi_b$ state $\Xi_{b}\ket{J^{P}=\frac{5}{2}^{-},2}_{\rho}$.}
	\centering
	\begin{tabular}{ccccccccc}	
		\hline\hline
		\multirow{2}{*}{Decay width}~~~~&$\underline{~~\Xi_{b}\ket{J^{P}=\frac{5}{2}^{-},2}_{\rho}~~}$ \\
&$\Xi_b(6227)^-$\\
\hline
$\Gamma[\Xi_b\pi]$&10.7\\
$\Gamma[\Lambda_bK]$&3.0\\
$\Gamma[\Xi_b'\pi]$&0.4\\
$\Gamma[\Xi_b^*\pi]$&0.4\\
$\Gamma_{\text{Total}}$&14.5\\ \hline
$\Gamma_{\text{Expt.}}$&$19.9\pm2.6$\\
\hline\hline
\end{tabular}
\end{table}

Our theoretical results indicate that the decay properties of the $\Xi_{b}\ket{J^{P}=\frac{5}{2}^{-},2}_{\rho}$ state is in good agreement with  the newly observed $\Xi_b(6227)^-$ in both the $\Lambda_b^-K$ and $\Xi_b^0\pi^-$ invariant mass spectra at LHCb~\cite{LHCb:2018vuc}. Fixing the mass at the physical mass $M=6228$ MeV, the total decay width of $\Xi_{b}\ket{J^{P}=\frac{5}{2}^{-},2}_{\rho}$ is about $\Gamma_{\text{Total}}\sim15$ MeV (see Table~\ref{Xib}), which is close to the lower limit of the measured width $\Gamma_{\text{Expt.}}\simeq19.9\pm2.6$ MeV. Furthermore, the dominant decay channels are $\Xi_b\pi$ and $\Lambda_bK$, which is consistent with the nature of $\Xi_b(6227)^-$.

Similarly, the predicted masses of the $\rho$-mode $1P$ $\Xi_c$ and $\Xi_b$ excitations certainly have a large uncertainty, which may bring uncertainties to the theoretical results. To investigate this effect, we plot the two-body strong decay widths of the $\rho$-mode $1P$ $\Xi_c$ and $\Xi_b$ excitations as a function of the mass in Fig.~\ref{fig-4}. As a whole, most of the $\rho$-mode $1P$ $\Xi_c$ and $\Xi_b$ states may have good potential to be observed in experiments due to their relatively narrow widths.

\subsection{$\Xi^{'}_c$ and $\Xi^{'}_b$ baryons }

In the $\Xi^{'}_c$ and $\Xi^{'}_{b}$ families, there are each two $\rho$-mode $1P$-wave excitations: $\ket{J^{P}=\frac{1}{2}^{-},1}_{\rho}$ and $\ket{J^{P}=\frac{3}{2}^{-},1}_{\rho}$. According to the theoretical predictions by various methods, the masses of the $\rho$-mode $1P$ $\Xi^{'}_c$ and $\Xi^{'}_{b}$ baryons are about $M\sim3.0$ and $M\sim6.3$ Gev, respectively. Fixing their masses at the predictions in Ref.~\cite{Bijker:2020tns}, we collect their strong decay properties in Table~\ref{Xiprime}.

\begin{table}[]
	\caption{\label{Xiprime}Partial and total strong decay widths of the $\rho$-mode $1P$ states in the $\Xi^{'}_{c}$ and $\Xi^{'}_{b}$ families. The unit of the width and mass is MeV. The masses for the unestablished $\rho$-mode $1P$ $\Xi^{'}_{c}$ and $\Xi^{'}_{b}$ states are taken from the predictions in Ref.~\cite{Bijker:2020tns}}
	\centering
	\begin{tabular}{c|ccccc}
		\hline \hline
		\multirow{2}{*}{States}&\multicolumn{5}{c}{Decay modes} \\
		\cline{2-6}
		&~~~$\Xi^{'}_c$$\pi$&~~~$\Xi^{*}_c$$\pi$&~~~$\Sigma_c$K&~~~$\Sigma^{*}_c$K&~~~Total\\
		\hline
		$\Xi^{'}_{c}\ket{J^{P}=\frac{1}{2}^{-},1}_{\rho}$(3060)&16.3&11.7&71.7&1.1&100.8\\
		\hline
		$\Xi^{'}_{c}\ket{J^{P}=\frac{3}{2}^{-},1}_{\rho}$(3096)&16.4&25.8&10.4&72.8&125.4\\
		\hline\hline
		\multirow{2}{*}{States}&\multicolumn{5}{c}{Decay modes} \\
		\cline{2-6}	
		&$\Xi^{'}_b$$\pi$&$\Xi^{*}_b$$\pi$&$\Sigma_b$K&$\Sigma^{*}_b$K&Total\\
		\hline
		$\Xi^{'}_{b}\ket{J^{P}=\frac{1}{2}^{-},1}_{\rho}$(6356)&16.9&13.4&69.6&0.6&100.4\\
		\hline	
		$\Xi^{'}_{b}\ket{J^{P}=\frac{3}{2}^{-},1}_{\rho}$(6364)&8.3&24.3&1.5&65.1&99.1\\
		\hline\hline	
	\end{tabular}
\end{table}

\begin{figure}[]
	\centering
	\subfigure[$1P_{\rho}$-wave $\Xi^{'}_c$ states]
	{\includegraphics[height=7cm,width=4.2cm]{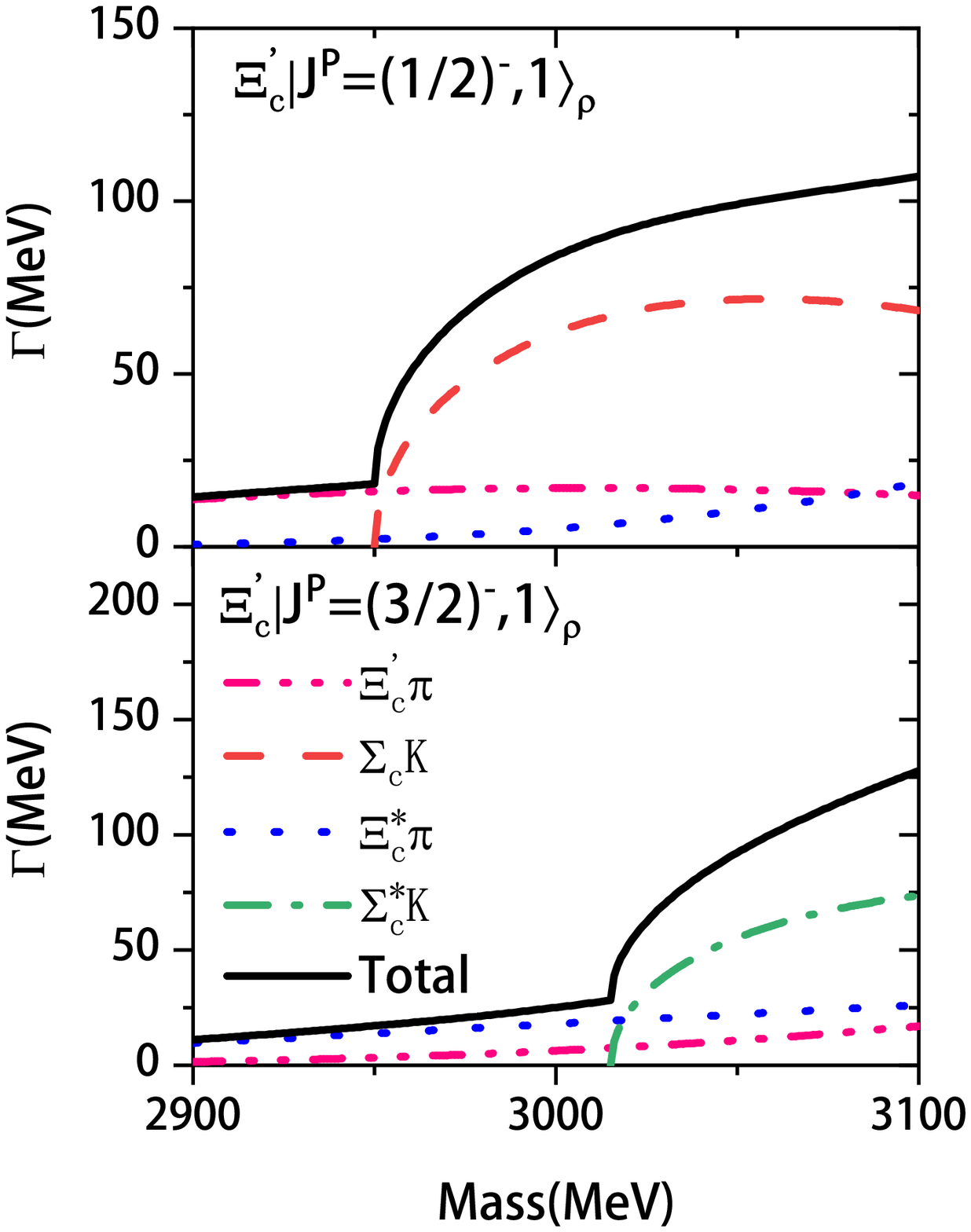}}
	\subfigure[$1P_{\rho}$-wave $\Xi^{'}_b$ states]
	{\includegraphics[height=7cm,width=4.0cm]{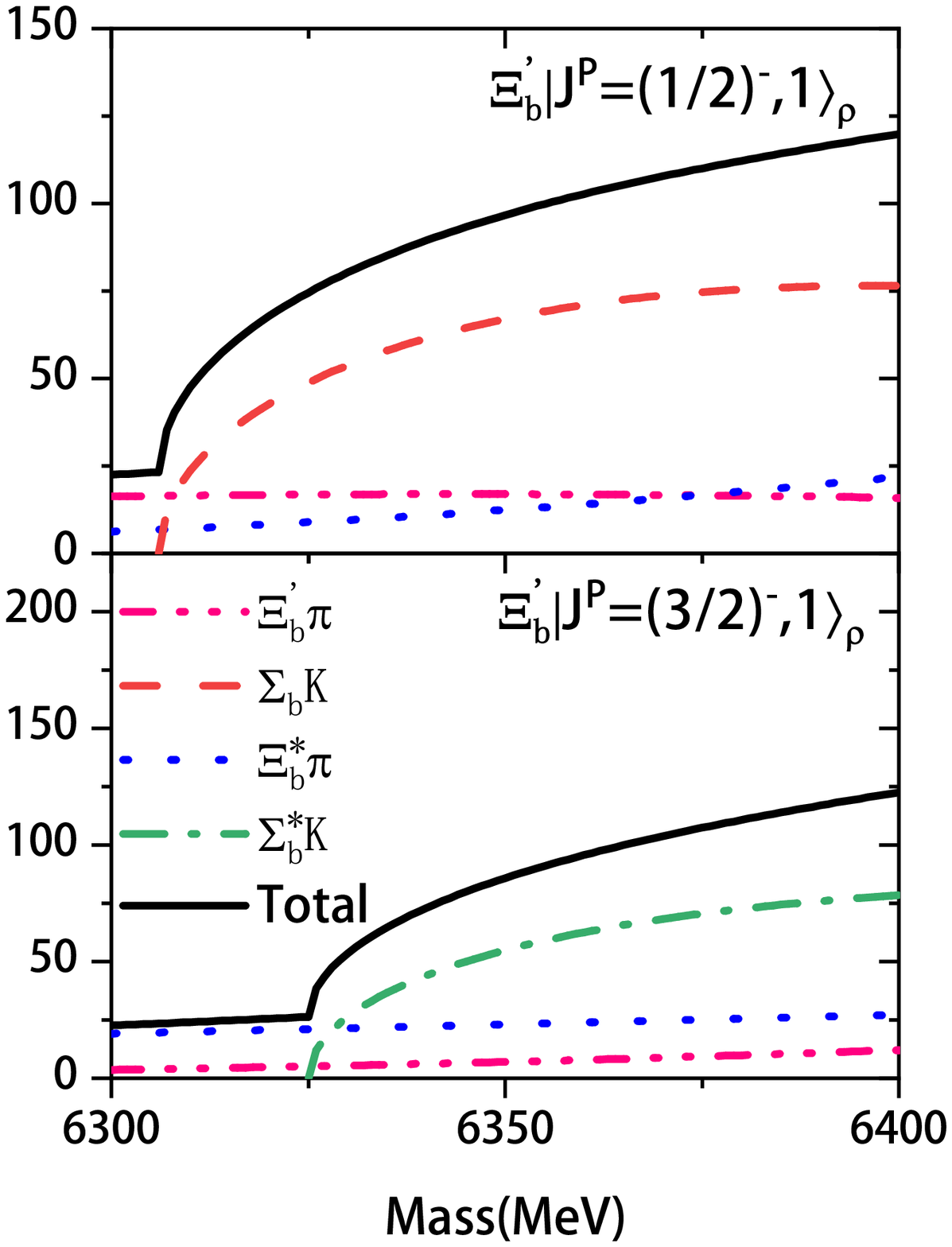}}
	\caption{Partial and total strong decay widths of the $\rho$-mode $1P$ states in the $\Xi^{'}_{c}$ and $\Xi^{'}_{b}$ families as functions of their masses.}\label{fig-5}
\end{figure}

All of the four states may be moderate states with a comparable width of $\Gamma\sim100$ MeV. While, their main decay channels have some difference.
The $\Xi^{'}_{c}\ket{J^{P}=\frac{1}{2}^{-},1}_{\rho}$ is governed by $\Sigma_cK$, and the corresponding branching fraction is predicted to be
\begin{eqnarray}
\frac{\Gamma[\Xi^{'}_{c}\ket{J^{P}=\frac{1}{2}^{-},1}_{\rho}\rightarrow \Sigma_cK]}{\Gamma_{\text{Total}}}\sim71\%.
\end{eqnarray}
Thus, the $\Xi^{'}_{c}\ket{J^{P}=\frac{1}{2}^{-},1}_{\rho}$ may be observed in the $\Lambda_c\pi K$ final state via the decay chain $\Xi^{'}_{c}\ket{J^{P}=\frac{1}{2}^{-},1}_{\rho}\rightarrow \Sigma_cK\rightarrow \Lambda_c\pi K$ in future experiments.
Meanwhile, this state has sizeable decay decay rates into $\Xi_c'\pi$ and $\Xi_c^*\pi$ with branching fractions
\begin{eqnarray}
\frac{\Gamma[\Xi^{'}_{c}\ket{J^{P}=\frac{1}{2}^{-},1}_{\rho}\rightarrow \Xi_c'\pi]}{\Gamma_{\text{Total}}}\sim16\%,\\
\frac{\Gamma[\Xi^{'}_{c}\ket{J^{P}=\frac{1}{2}^{-},1}_{\rho}\rightarrow \Xi_c^*\pi]}{\Gamma_{\text{Total}}}\sim12\%.
\end{eqnarray}

For the $\Xi^{'}_{c}\ket{J^{P}=\frac{3}{2}^{-},1}_{\rho}$, it mainly decays into $\Sigma_c^*K$ with a branching fraction
\begin{eqnarray}
\frac{\Gamma[\Xi^{'}_{c}\ket{J^{P}=\frac{3}{2}^{-},1}_{\rho}\rightarrow \Sigma_c^*K]}{\Gamma_{\text{Total}}}\sim58\%.
\end{eqnarray}
Yet the other channels $\Xi_c'\pi$, $\Xi_c^*\pi$ and $\Sigma_cK$ are not obviously neglectable as well. The branching fraction for the $\Xi_c^*\pi$
channel can reach up to
\begin{eqnarray}
\frac{\Gamma[\Xi^{'}_{c}\ket{J^{P}=\frac{3}{2}^{-},1}_{\rho}\rightarrow \Xi_c^*\pi]}{\Gamma_{\text{Total}}}\sim21\%.
\end{eqnarray}

For the two $\rho$-mode $1P$ states in the $\Xi_b'$ family, the dominant decay channel of $\Xi^{'}_{b}\ket{J^{P}=\frac{1}{2}^{-},1}_{\rho}$ is $\Sigma_bK$, while that of $\Xi^{'}_{b}\ket{J^{P}=\frac{3}{2}^{-},1}_{\rho}$ is $\Sigma_b^*K$. The predicted branching fractions are
\begin{eqnarray}
\frac{\Gamma[\Xi^{'}_{b}\ket{J^{P}=\frac{1}{2}^{-},1}_{\rho}\rightarrow \Sigma_bK]}{\Gamma_{\text{Total}}}\sim69\%,\\
\frac{\Gamma[\Xi^{'}_{b}\ket{J^{P}=\frac{3}{2}^{-},1}_{\rho}\rightarrow \Sigma_b^*K]}{\Gamma_{\text{Total}}}\sim66\%.
\end{eqnarray}
In addition, the $\Xi^{'}_{b}\ket{J^{P}=\frac{3}{2}^{-},1}_{\rho}$ has a large decay rate into $\Xi_b^*\pi$ with the branching fraction
\begin{eqnarray}
\frac{\Gamma[\Xi^{'}_{b}\ket{J^{P}=\frac{3}{2}^{-},1}_{\rho}\rightarrow \Xi_b^*\pi]}{\Gamma_{\text{Total}}}\sim25\%.
\end{eqnarray}

We also plot the partial decay widths of the $\rho$-mode $1P$ $\Xi_c'$ and $\Xi_b'$ baryons as a function of the mass in Fig.~\ref{fig-5}. The two-body strong decays of the $\rho$-mode $1P$ state in $\Xi_c'$ family are similar to that in $\Xi_b'$ family. Roughly speaking, they are only  containing different heavy quarks: In the $\Xi_c'$ family the heavy quark is $c$ quark, while in $\Xi_b'$ family the heavy quark changes to $b$ quark.

\subsection{$\Omega_c$ and $\Omega_b$ baryons }

In the $\Omega_{c}$ and $\Omega_{b}$ families, there are also two $\rho$-mode $1P$-wave states,
$\ket{J^{P}=\frac{1}{2}^{-},1}$ and $\ket{J^{P}=\frac{3}{2}^{-},1}$. The masses of these $\rho$-mode $1P$-wave $\Omega_c$ and $\Omega_{b}$ states are respectively about $M\sim$ 3.1 and $M\sim$ 6.4 GeV within various quark model predictions. Similarly, we first fix their masses at the predictions in Ref.~\cite{Yoshida:2015tia}, and collect the decay properties in Table~\ref{Omega}.

\begin{table}[]
\caption{\label{Omega}Partial and total strong decay widths of the $\rho$-mode $1P$ states in the  $\Omega_{c}$ and $\Omega_{b}$ families. The unit of the width and mass is MeV. The masses for the unestablished $\rho$-mode $1P$ $\Omega_{c}$ and $\Omega_{b}$ states are taken from the predictions in Ref.~\cite{Yoshida:2015tia}.}
\centering
\begin{tabular}{c|ccc}
	\hline\hline
	\multirow{2}{*}{States}~~~&~~~~~~~\underline{$\Omega_{c}\ket{J^{P}=\frac{1}{2}^{-},1}_{\rho}$} ~~~~~~~ &\underline{$\Omega_{c}\ket{J^{P}=\frac{3}{2}^{-},1}_{\rho}$}\\
	&$M=3110$~~~~~~~&$M=3112$\\
\hline
    $\Gamma[\Xi^{'}_cK] $~~~ &119.4~~~~~~~&0.9\\
  \hline\hline
  \multirow{2}{*}{States}~~~&~~~~~~~\underline{$\Omega_{b}\ket{J^{P}=\frac{1}{2}^{-},1}_{\rho}$} ~~~~~~~&\underline{$\Omega_{b}\ket{J^{P}=\frac{3}{2}^{-},1}_{\rho}$}\\
&$M=6437$~~~~~~~&$M=6438$\\	
\hline
    $\Gamma[\Xi^{'}_bK]$~~~  &65.2~~~~~~~&0.02\\
		\hline\hline
	\end{tabular}
\end{table}

In the $\Omega_c$ family, both the two $\rho$-mode $1P$ states mainly decay into $\Xi_c'K$. However, the $J^P=1/2^-$ state $\Omega_{c}\ket{J^{P}=\frac{1}{2}^{-},1}_{\rho}$ may be a moderate state with a width of $\Gamma\sim119$ MeV, while the $J^P=3/2^-$ state $\Omega_{c}\ket{J^{P}=\frac{3}{2}^{-},1}_{\rho}$ is most likely to be a very narrow state with a width of $\Gamma\sim0.9$ MeV.

\begin{figure}[]
	\centering
	\subfigure[$1P_{\rho}$-wave $\Omega_c$ states]
	{\includegraphics[height=7cm,width=4.2cm]{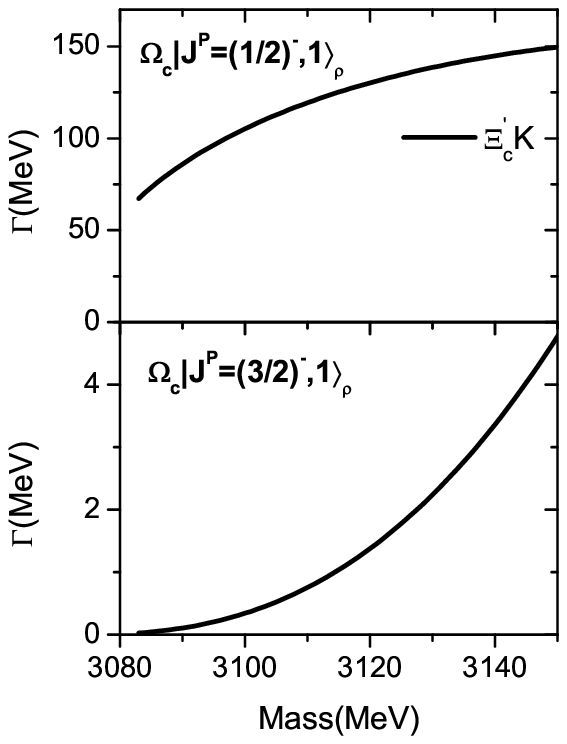}}
	\subfigure[$1P_{\rho}$-wave $\Omega_b$ states]
	{\includegraphics[height=7cm,width=4.0cm]{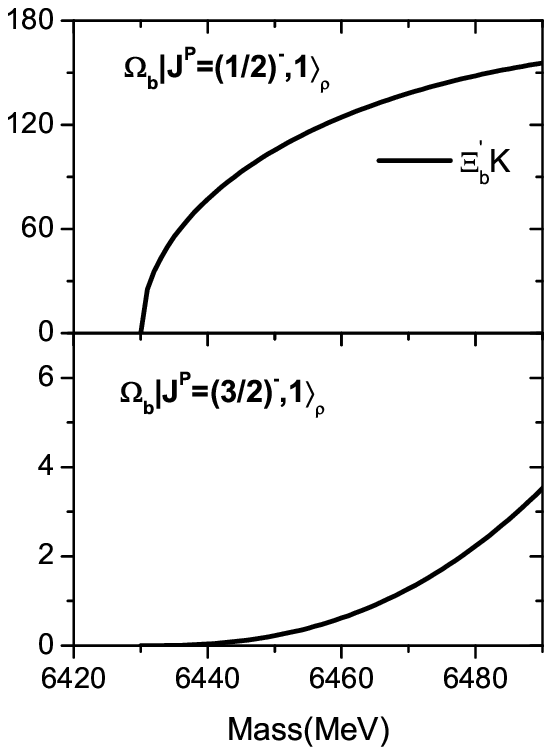}}
	\caption{Partial and total strong decay width of the $\rho$-mode 1P states in the $\Omega_{c}$ and $\Omega_{b}$ families as functions of their mass.}\label{fig-6}
\end{figure}


For the two $\rho$-mode $1P$ $\Omega_b$ states, their decays are governed by $\Xi_b'K$. The $J^P=1/2^-$ state $\Omega_{b}\ket{J^{P}=\frac{1}{2}^{-},1}_{\rho}$ has a width of $\Gamma\sim65$ MeV, and may be observed in the $\Xi_b\pi K$ final state via the decay chain $\Omega_{b}\ket{J^{P}=\frac{1}{2}^{-},1}_{\rho}\rightarrow \Xi_b'K\rightarrow\Xi_b\pi K$. The $J^P=3/2^-$ state $\Omega_{b}\ket{J^{P}=\frac{3}{2}^{-},1}_{\rho}$ may a particularly narrow state with a width of $\Gamma<0.1$ MeV. Thus, the state $\Omega_{b}\ket{J^{P}=\frac{3}{2}^{-},1}_{\rho}$ may have a large potential to be observed in the $\Xi_b\pi K$ final state via the decay chain $\Omega_{b}\ket{J^{P}=\frac{3}{2}^{-},1}_{\rho}\rightarrow \Xi_b'K\rightarrow\Xi_b\pi K$ at LHCb.

In addition, we analyze the decay properties of the $\rho$-mode $1P$ $\Omega_c$ and $\Omega_b$ baryons as a function of the mass in Fig.~\ref{fig-6}. It should be pointed out that the total width of the $J^P=3/2^-$ state $\Omega_{c(b)}\ket{J^{P}=\frac{3}{2}^{-},1}_{\rho}$ is $\Gamma<5$ MeV within the mass varying in the region of what we considered , which indicates the state $\Omega_{c(b)}\ket{J^{P}=\frac{3}{2}^{-},1}_{\rho}$ has a good potential to be observed in future experiments.


\section{Summary}

In the present work, we systematically studied the two-body strong decays of the low-lying $\rho$-mode $1P$-wave singly heavy baryons in chiral quark model within the $j$-$j$ coupling scheme. On the one hand, we attempt to confirm the possibility of the controversial singly heavy baryons taken as $\rho$-mode excitations. On the other hand, we hope to provide the theories foundation for the experiment exploring of the missing $\rho$-mode $1P$ singly heavy baryons. Our main results are summarized as follows.

For the $\rho$-mode $1P$ $\Lambda_c$ and $\Lambda_b$ baryons, the OZI-allowed two-body strong decays of the $J^P=1/2^-$ state $\Lambda_{c(b)}\ket{J^{P}=\frac{1}{2}^{-},0}_{\rho}$ are forbidden since we adopt the simple harmonic oscillator wave functions. Hence, the $\Lambda_{c(b)}\ket{J^{P}=\frac{1}{2}^{-},0}_{\rho}$ should be very narrow state. The three $\rho$-mode states $\Lambda_{c(b)}\ket{J^{P}=\frac{1}{2}^{-},1}_{\rho}$, $\Lambda_{c(b)}\ket{J^{P}=\frac{3}{2}^{-},1}_{\rho}$ and $ \Lambda_{c(b)}\ket{J^{P}=\frac{3}{2}^{-},2}_{\rho}$ have a broad width of $\Gamma\sim(100-200)$ MeV. It can be a big challenge for experimenters to observe them in future. While, the $J^P=5/2^-$ state $\Lambda_{c(b)}\ket{J^{P}=\frac{5}{2}^{-},2}_{\rho}$ may be a narrow state and the decay width is about dozens of MeV. Thus, the $\Lambda_{c}\ket{J^{P}=\frac{5}{2}^{-},2}_{\rho}$ has a good potential to be observed in the $\Lambda_{c}\pi\pi$ final state via the decay chain $\Lambda_{c}\ket{J^{P}=\frac{5}{2}^{-},2}_{\rho}\rightarrow \Sigma_{c}\pi\rightarrow\Lambda_{c}\pi\pi$, and the $\Lambda_b\ket{J^{P}=\frac{5}{2}^{-},2}_{\rho}$ may be observed in the $\Lambda_{b}\pi\pi$ final state via the decay chain $\Lambda_{b}\ket{J^{P}=\frac{5}{2}^{-},2}_{\rho}\rightarrow \Sigma_{b}\pi/\Sigma^*_{b}\pi\rightarrow\Lambda_{b}\pi\pi$.
The newly observed state $\Lambda_c(2910)^+$ can be explained as the $J^P=5/2^-$ state $\Lambda_{c}\ket{J^{P}=\frac{5}{2}^{-},2}_{\rho}$.


The $\rho$-mode $1P$ $\Sigma_{c(b)}$ and $\Xi'_{c(b)}$ states are predicted to be broad states with a decay width of around $\Gamma\sim(150-200)$ and $\Gamma\sim100$ MeV, respectively. It should be mentioned that if the mass of the $J^P=3/2^-$ state $\Xi'_{c(b)}\ket{J^{P}=\frac{3}{2}^{-},1}_{\rho}$ is close to the mass threshold of the $\Sigma^*_{c(b)}K$, this state may be a narrow state with dozens of MeV. Hence, the $\Xi'_{c(b)}\ket{J^{P}=\frac{3}{2}^{-},1}_{\rho}$ may be observed via its mainly decay channel $\Xi_{c(b)}\pi$.

For the $\rho$-mode $1P$ $\Xi_c$ and $\Xi_b$ baryons, the states $\Xi_{b}\ket{J^{P}=\frac{1}{2}^{-},1}_{\rho}$,  $ \Xi_{c(b)}\ket{J^{P}=\frac{3}{2}^{-},2}_{\rho} $,  $\Xi_{c(b)} \ket{J^{P}=\frac{3}{2}^{-},2}_{\rho} $ and $\Xi_{c(b)} \ket{J^{P}=\frac{5}{2}^{-},2}_{\rho}$ may be narrow states with a total decay width around dozens of MeV, and have a good potential to be observed in experiments, especially the $\Xi_{b}$ states, whose decay widths are relatively narrower. The ideal channels for exploring these missing $\rho$-mode $1P$ $\Xi_{c}$ states may be $\Xi_c$$\pi$, $\Xi^{*}_{c}$$\pi$ and $\Lambda_cK$, while that for the $\Xi_b$ states may be $\Xi_{b}$$\pi$, $\Xi^{'}_{b}\pi$ and $\Xi^{*}_{b}\pi$. Furthermore, our results indicate that the decay properties of the $\Xi_{b} \ket{J^{P}=\frac{5}{2}^{-},2}_{\rho}$ state is in good agreement with the newly observed state $\Xi_b(6227)^-$, which mainly decays into $\Xi_b\pi$ and $\Lambda_bK$.

As to the $\rho$-mode $1P$ $\Omega_c$ and $\Omega_b$ baryons, the $\Omega_{c(b)}\ket{J^{P}=\frac{3}{2}^{-},1} $ state also have good potentials to be observed in experiments due to its particularly narrow width of a few MeV.  To looking for these sates, the $\Xi^{'}_{c(b)}K$ is worth observing in future experiments.

\section*{Acknowledgements }

This work is supported by the National Natural Science Foundation of China under Grants No.12005013, No.11947048, No.12175065, No.U1832173 and No.11775078.

 \bibliography{documents}

\begin{thebibliography}{10}

\bibitem{Grozin:1992yq}
A.~G. Grozin.
\newblock {Introduction to the heavy quark effective theory. part 1}.
\newblock 12 1992.

\bibitem{Mannel:1996rg}
T.~Mannel.
\newblock {Effective theory for heavy quarks}.
\newblock {\em Lect. Notes Phys.}, 479:387--428, 1997.

\bibitem{ParticleDataGroup:2020ssz}
P.~A. Zyla et~al.
\newblock {Review of Particle Physics}.
\newblock {\em PTEP}, 2020(8):083C01, 2020.

\bibitem{LHCb:2020iby}
Roel Aaij et~al.
\newblock {Observation of New $\Xi_c^0$ Baryons Decaying to $\Lambda_c^+ K^-$}.
\newblock {\em Phys. Rev. Lett.}, 124(22):222001, 2020.

\bibitem{Belle:2022hnm}
{Evidence of a new excited charmed baryon decaying to $\Sigma_{c}(2455)^{0,++}
  \pi^{\pm}$}.
\newblock 6 2022.

\bibitem{LHCb:2021ssn}
Roel Aaij et~al.
\newblock {Observation of Two New Excited $\Xi_b^0$ States Decaying to
  $\Lambda^0_b K^- \pi^+$}.
\newblock {\em Phys. Rev. Lett.}, 128(16):162001, 2022.

\bibitem{Yoshida:2015tia}
Tetsuya Yoshida, Emiko Hiyama, Atsushi Hosaka, Makoto Oka, and Katsunori
  Sadato.
\newblock {Spectrum of heavy baryons in the quark model}.
\newblock {\em Phys. Rev. D}, 92(11):114029, 2015.

\bibitem{Lu:2018utx}
Qi-Fang L\"u, Li-Ye Xiao, Zuo-Yun Wang, and Xian-Hui Zhong.
\newblock {Strong decay of $\Lambda _c(2940)$ as a $2P$ state in the $\Lambda
  _c$ family}.
\newblock {\em Eur. Phys. J. C}, 78(7):599, 2018.

\bibitem{Wang:2017hej}
Kai-Lei Wang, Li-Ye Xiao, Xian-Hui Zhong, and Qiang Zhao.
\newblock {Understanding the newly observed $\Omega_c$ states through their
  decays}.
\newblock {\em Phys. Rev. D}, 95(11):116010, 2017.

\bibitem{Liu:2012sj}
Lei-Hua Liu, Li-Ye Xiao, and Xian-Hui Zhong.
\newblock {Charm-strange baryon strong decays in a chiral quark model}.
\newblock {\em Phys. Rev. D}, 86:034024, 2012.

\bibitem{Guo:2019ytq}
Jing-Jing Guo, Pei Yang, and Ailin Zhang.
\newblock {Strong decays of observed $\Lambda_c$ baryons in the $^3P_0$ model}.
\newblock {\em Phys. Rev. D}, 100(1):014001, 2019.

\bibitem{Chen:2007xf}
Chong Chen, Xiao-Lin Chen, Xiang Liu, Wei-Zhen Deng, and Shi-Lin Zhu.
\newblock {Strong decays of charmed baryons}.
\newblock {\em Phys. Rev. D}, 75:094017, 2007.

\bibitem{Chen:2015kpa}
Hua-Xing Chen, Wei Chen, Qiang Mao, Atsushi Hosaka, Xiang Liu, and Shi-Lin Zhu.
\newblock {P-wave charmed baryons from QCD sum rules}.
\newblock {\em Phys. Rev. D}, 91(5):054034, 2015.

\bibitem{Chen:2016iyi}
Bing Chen, Ke-Wei Wei, Xiang Liu, and Takayuki Matsuki.
\newblock {Low-lying charmed and charmed-strange baryon states}.
\newblock {\em Eur. Phys. J. C}, 77(3):154, 2017.

\bibitem{Cheng:2017ove}
Hai-Yang Cheng and Cheng-Wei Chiang.
\newblock {Quantum numbers of $\Omega_c$ states and other charmed baryons}.
\newblock {\em Phys. Rev. D}, 95(9):094018, 2017.

\bibitem{Luo:2019qkm}
Si-Qiang Luo, Bing Chen, Zhan-Wei Liu, and Xiang Liu.
\newblock {Resolving the low mass puzzle of $\Lambda_c(2940)^+$}.
\newblock {\em Eur. Phys. J. C}, 80(4):301, 2020.

\bibitem{Chen:2018orb}
Bing Chen, Ke-Wei Wei, Xiang Liu, and Ailin Zhang.
\newblock {Role of newly discovered $\Xi_b(6227)^-$ for constructing excited
  bottom baryon family}.
\newblock {\em Phys. Rev. D}, 98(3):031502, 2018.

\bibitem{Chen:2017gnu}
Bing Chen and Xiang Liu.
\newblock {New $\Omega_c^0$ baryons discovered by LHCb as the members of $1P$
  and $2S$ states}.
\newblock {\em Phys. Rev. D}, 96(9):094015, 2017.

\bibitem{Yang:2021lce}
Hui-Min Yang and Hua-Xing Chen.
\newblock {$P$-wave charmed baryons of the $SU(3)$ flavor $6_F$}.
\newblock {\em Phys. Rev. D}, 104(3):034037, 2021.

\bibitem{Roberts:2007ni}
W.~Roberts and Muslema Pervin.
\newblock {Heavy baryons in a quark model}.
\newblock {\em Int. J. Mod. Phys. A}, 23:2817--2860, 2008.

\bibitem{Grach:2008ij}
I.~L. Grach, I.~M. Narodetskii, M.~A. Trusov, and A.~I. Veselov.
\newblock {Heavy baryon spectroscopy in the QCD string model}.
\newblock In {\em {18th International Conference on Particles and Nuclei}}, 11
  2008.

\bibitem{Bijker:2020tns}
Roelof Bijker, Hugo Garc\'\i{}a-Tecocoatzi, Alessandro Giachino, Emmanuel
  Ortiz-Pacheco, and Elena Santopinto.
\newblock {Masses and decay widths of $\Xi_{c/b}$ and $\Xi^\prime_{c/b}$
  baryons}.
\newblock 10 2020.

\bibitem{Zhong:2007gp}
Xian-Hui Zhong and Qiang Zhao.
\newblock {Charmed baryon strong decays in a chiral quark model}.
\newblock {\em Phys. Rev. D}, 77:074008, 2008.

\bibitem{Wang:2021bmz}
Kai-Lei Wang and Xian-Hui Zhong.
\newblock {Toward establishing the low-lying $P$-wave excited $\Sigma_c$ baryon
  states}.
\newblock {\em Chin. Phys. C}, 46:2, 2022.

\bibitem{Xiao:2020gjo}
Li-Ye Xiao and Xian-Hui Zhong.
\newblock {Toward establishing the low-lying $P$-wave $\Sigma_b$ states}.
\newblock {\em Phys. Rev. D}, 102(1):014009, 2020.

\bibitem{Wang:2020gkn}
Kai-Lei Wang, Li-Ye Xiao, and Xian-Hui Zhong.
\newblock {Understanding the newly observed $\Xi_c^0$ states through their
  decays}.
\newblock {\em Phys. Rev. D}, 102(3):034029, 2020.

\bibitem{Xiao:2020oif}
Li-Ye Xiao, Kai-Lei Wang, Ming-Sheng Liu, and Xian-Hui Zhong.
\newblock {Possible interpretation of the newly observed $\Omega _b$ states}.
\newblock {\em Eur. Phys. J. C}, 80(3):279, 2020.

\bibitem{Wang:2018fjm}
Kai-Lei Wang, Qi-Fang L\"u, and Xian-Hui Zhong.
\newblock {Interpretation of the newly observed $\Sigma_b(6097)^{\pm}$ and
  $\Xi_b(6227)^-$ states as the $P$-wave bottom baryons}.
\newblock {\em Phys. Rev. D}, 99(1):014011, 2019.

\bibitem{Wang:2019uaj}
Kai-Lei Wang, Qi-Fang L\"u, and Xian-Hui Zhong.
\newblock {Interpretation of the newly observed $\Lambda_b(6146)^{0}$ and
  $\Lambda_b(6152)^0$ states in a chiral quark model}.
\newblock {\em Phys. Rev. D}, 100(11):114035, 2019.

\bibitem{Wang:2022zqv}
Wen-Jia Wang, Yu-Hui Zhou, Li-Ye Xiao, and Xian-Hui Zhong.
\newblock {The $1D$-wave bottom-strange baryons and possible interpretation of
  $\Xi_{b}(6327)^{0}$ and $\Xi_{b}(6333)^{0}$}.
\newblock 2 2022.

\bibitem{Lu:2020ivo}
Qi-Fang L\"u.
\newblock {Canonical interpretations of the newly observed $\Xi _c(2923)^0$,
  $\Xi _c(2939)^0$, and $\Xi _c(2965)^0$ resonances}.
\newblock {\em Eur. Phys. J. C}, 80(10):921, 2020.

\bibitem{Kakadiya:2022zvy}
Amee Kakadiya, Zalak Shah, and Ajay~Kumar Rai.
\newblock {Mass Spectra and Decay Properties of Singly Heavy Bottom-Strange
  Baryons}.
\newblock 2 2022.

\bibitem{Zhao:2020tpf}
Ze~Zhao.
\newblock {Theoretical interpretation of $\Xi_c(2970)$}.
\newblock {\em Phys. Rev. D}, 102(9):096021, 2020.

\bibitem{Suenaga:2022ajn}
Daiki Suenaga and Atsushi Hosaka.
\newblock {Decays of Roper-like singly heavy baryons in a chiral model}.
\newblock 2 2022.

\bibitem{Azizi:2020azq}
K.~Azizi, Y.~Sarac, and H.~Sundu.
\newblock {Determination of the possible quantum numbers for the newly observed
  $\Xi_b(6227)^0$ state}.
\newblock {\em JHEP}, 03:244, 2021.

\bibitem{Yu:2022ymb}
Guo-Liang Yu, Zhen-Yu Li, Zhi-Gang Wang, Jie Lu, and Meng Yan.
\newblock {Systematic analysis of single heavy baryons $\Lambda_{Q}$,
  $\Sigma_{Q}$ and $\Omega_{Q}$}.
\newblock 6 2022.

\bibitem{Yu:2021zvl}
Guo-Liang Yu, Zhi-Gang Wang, and Xiu-Wu Wang.
\newblock {The $1D$, $2D$ $\Xi_{b}$ and $\Lambda_{b}$ baryons}.
\newblock 9 2021.

\bibitem{Yang:2020zjl}
Hui-Min Yang, Hua-Xing Chen, and Qiang Mao.
\newblock {Identifying the $\Xi_c^0$ baryons observed by LHCb as $P$-wave
  $\Xi_c^\prime$ baryons}.
\newblock {\em Phys. Rev. D}, 102:114009, 2020.

\bibitem{Wang:2017vnc}
Wei Wang and Rui-Lin Zhu.
\newblock {Interpretation of the newly observed $\Omega_c^0$ resonances}.
\newblock {\em Phys. Rev. D}, 96(1):014024, 2017.

\bibitem{Padmanath:2017lng}
M.~Padmanath and Nilmani Mathur.
\newblock {Quantum Numbers of Recently Discovered $\Omega^{0}_{c}$ Baryons from
  Lattice QCD}.
\newblock {\em Phys. Rev. Lett.}, 119(4):042001, 2017.

\bibitem{Karliner:2017kfm}
Marek Karliner and Jonathan~L. Rosner.
\newblock {Very narrow excited $\Omega_c$ baryons}.
\newblock {\em Phys. Rev. D}, 95(11):114012, 2017.

\bibitem{Wang:2017zjw}
Zhi-Gang Wang.
\newblock {Analysis of $\Omega _c(3000)$ , $\Omega _c(3050)$ , $\Omega
  _c(3066)$ , $\Omega _c(3090)$ and $\Omega _c(3119)$ with QCD sum rules}.
\newblock {\em Eur. Phys. J. C}, 77(5):325, 2017.

\bibitem{Yang:2019cvw}
Hui-Min Yang, Hua-Xing Chen, Er-Liang Cui, Atsushi Hosaka, and Qiang Mao.
\newblock {Decay properties of $P$-wave bottom baryons within light-cone sum
  rules}.
\newblock {\em Eur. Phys. J. C}, 80(2):80, 2020.

\bibitem{Cui:2019dzj}
Er-Liang Cui, Hui-Min Yang, Hua-Xing Chen, and Atsushi Hosaka.
\newblock {Identifying the $\Xi_{b}(6227)$ and $\Sigma_{b}(6097)$ as $P$-wave
  bottom baryons of $J^P = 3/2^-$}.
\newblock {\em Phys. Rev. D}, 99(9):094021, 2019.

\bibitem{Yang:2020zrh}
Hui-Min Yang and Hua-Xing Chen.
\newblock {$P$-wave bottom baryons of the $SU(3)$ flavor $\mathbf{6}_F$}.
\newblock {\em Phys. Rev. D}, 101(11):114013, 2020.
\newblock [Erratum: Phys.Rev.D 102, 079901 (2020)].

\bibitem{Aliev:2018vye}
T.~M. Aliev, K.~Azizi, Y.~Sarac, and H.~Sundu.
\newblock {Determination of the quantum numbers of $\Sigma_b(6097)^{\pm}$ via
  their strong decays}.
\newblock {\em Phys. Rev. D}, 99(9):094003, 2019.

\bibitem{Chen:2018vuc}
Bing Chen and Xiang Liu.
\newblock {Assigning the newly reported $\Sigma_b(6097)$ as a $P$-wave excited
  state and predicting its partners}.
\newblock {\em Phys. Rev. D}, 98(7):074032, 2018.

\bibitem{Yang:2018lzg}
Pei Yang, Jing-Jing Guo, and Ailin Zhang.
\newblock {Identification of the newly observed $\Sigma_b(6097)^\pm$ baryons
  from their strong decays}.
\newblock {\em Phys. Rev. D}, 99(3):034018, 2019.

\bibitem{Jia:2019bkr}
Duojie Jia, Wen-Nian Liu, and Atsushi Hosaka.
\newblock {Regge behaviors in orbitally excited spectroscopy of charmed and
  bottom baryons}.
\newblock {\em Phys. Rev. D}, 101(3):034016, 2020.

\bibitem{He:2021xrh}
Hui-Zhen He, Wei Liang, Qi-Fang L\"u, and Yu-Bing Dong.
\newblock {Strong decays of the low-lying bottom strange baryons}.
\newblock {\em Sci. China Phys. Mech. Astron.}, 64(6):261012, 2021.

\bibitem{Wang:2020pri}
Zhi-Gang Wang.
\newblock {Analysis of the $\Omega_b(6316)$, $\Omega_b(6330)$, $\Omega_b(6340)$
  and $\Omega_b(6350)$ with QCD sum rules}.
\newblock {\em Int. J. Mod. Phys. A}, 35(07):2050043, 2020.

\bibitem{Mutuk:2020rzm}
Halil Mutuk.
\newblock {A study of excited $\Omega _b^-$ states in hypercentral constituent
  quark model via artificial neural network}.
\newblock {\em Eur. Phys. J. A}, 56(5):146, 2020.

\bibitem{Liang:2020hbo}
Wei Liang and Qi-Fang L\"u.
\newblock {Strong decays of the newly observed narrow $\Omega _b$ structures}.
\newblock {\em Eur. Phys. J. C}, 80(3):198, 2020.

\bibitem{Azizi:2022dpn}
K.~Azizi, Y.~Sarac, and H.~Sundu.
\newblock {Interpretation of the $\Lambda_c(2910)^+$ baryon newly seen by Belle
  Collaboration and its possible bottom partner}.
\newblock 7 2022.

\bibitem{Agaev:2017lip}
S.~S. Agaev, K.~Azizi, and H.~Sundu.
\newblock {Interpretation of the new $\Omega_c^{0}$ states via their mass and
  width}.
\newblock {\em Eur. Phys. J. C}, 77(6):395, 2017.

\bibitem{Agaev:2020fut}
S.~S. Agaev, K.~Azizi, and H.~Sundu.
\newblock {Newly discovered $\Xi _c^{0}$ resonances and their parameters}.
\newblock {\em Eur. Phys. J. A}, 57(6):201, 2021.

\bibitem{Agaev:2017jyt}
S.~S. Agaev, K.~Azizi, and H.~Sundu.
\newblock {On the nature of the newly discovered $\Omega$ states}.
\newblock {\em EPL}, 118(6):61001, 2017.

\bibitem{Huang:2017dwn}
Hongxia Huang, Jialun Ping, and Fan Wang.
\newblock {Investigating the excited $\Omega^{0}_{c}$ states through $\Xi_{c}K$
  and $\Xi^{'}_{c}K$ decay channels}.
\newblock {\em Phys. Rev. D}, 97(3):034027, 2018.

\bibitem{Liu:2018bkx}
Ming-Zhu Liu, Tian-Wei Wu, Ju-Jun Xie, Manuel Pavon~Valderrama, and Li-Sheng
  Geng.
\newblock {$D \Xi$ and $D^* \Xi$ molecular states from one boson exchange}.
\newblock {\em Phys. Rev. D}, 98(1):014014, 2018.

\bibitem{Kishore:2019fzb}
Raj Kishore, Asmita Mukherjee, and Sangem Rajesh.
\newblock {Sivers asymmetry in the photoproduction of a $J/\psi$ and a jet at
  the EIC}.
\newblock {\em Phys. Rev. D}, 101(5):054003, 2020.

\bibitem{Huang:2018bed}
Yin Huang, Cheng-jian Xiao, Li-Sheng Geng, and Jun He.
\newblock {Strong decays of the $\Xi_b(6227)$ as a $\Sigma_b\bar{K}$ molecule}.
\newblock {\em Phys. Rev. D}, 99(1):014008, 2019.

\bibitem{Wang:2021cku}
Hui-Juan Wang, Zun-Yan Di, and Zhi-Gang Wang.
\newblock {Analysis of the excited \ensuremath{\Omega} c states as the
  \ensuremath{\pm} pentaquark states with QCD sum rules}.
\newblock {\em Commun. Theor. Phys.}, 73(3):035201, 2021.

\bibitem{Hu:2020zwc}
Xiaohuang Hu, Yue Tan, and Jialun Ping.
\newblock {Investigation of $\Xi_c^0$ in a chiral quark model}.
\newblock {\em Eur. Phys. J. C}, 81(4):370, 2021.

\bibitem{Yu:2018yxl}
Q.~X. Yu, R.~Pavao, V.~R. Debastiani, and E.~Oset.
\newblock {Description of the $\Xi _c$ and $\Xi _b$ states as molecular
  states}.
\newblock {\em Eur. Phys. J. C}, 79(2):167, 2019.

\bibitem{Wang:2020vwl}
Hui-Juan Wang, Zun-Yan Di, and Zhi-Gang Wang.
\newblock {Analysis of the \ensuremath{\Xi}$_{b}$(6227) as the $\frac
  {1}{2}^{\pm }$ Pentaquark Molecular States with QCD Sum Rules}.
\newblock {\em Int. J. Theor. Phys.}, 59(10):3124--3133, 2020.

\bibitem{Capstick:1986ter}
Simon Capstick and Nathan Isgur.
\newblock {Baryons in a relativized quark model with chromodynamics}.
\newblock {\em Phys. Rev. D}, 34(9):2809--2835, 1986.

\bibitem{Ozdem:2021vry}
Ula\c{s} \"Ozdem.
\newblock {Magnetic moment of the $\Xi _b(6227)$ as a molecular pentaquark
  state}.
\newblock {\em Eur. Phys. J. Plus}, 137(1):103, 2022.

\bibitem{Zhu:2020lza}
HongQiang Zhu and Yin Huang.
\newblock {Radiative decay of $\Xi_b(6227)$ in a hadronic molecule picture}.
\newblock {\em Chin. Phys. C}, 44(8):083101, 2020.

\bibitem{Chen:2021eyk}
Bing Chen, Si-Qiang Luo, and Xiang Liu.
\newblock {Universal behavior of mass gaps existing in the single heavy baryon
  family}.
\newblock {\em Eur. Phys. J. C}, 81(5):474, 2021.

\bibitem{Wang:2017kfr}
Kai-Lei Wang, Ya-Xiong Yao, Xian-Hui Zhong, and Qiang Zhao.
\newblock {Strong and radiative decays of the low-lying $S$- and $P$-wave
  singly heavy baryons}.
\newblock {\em Phys. Rev. D}, 96(11):116016, 2017.

\bibitem{Li:1994cy}
Zhen-Ping Li.
\newblock {The Threshold pion photoproduction of nucleons in the chiral quark
  model}.
\newblock {\em Phys. Rev. D}, 50:5639--5646, 1994.

\bibitem{Li:1997gd}
Zhen-ping Li, Hong-xing Ye, and Ming-hui Lu.
\newblock {An Unified approach to pseudoscalar meson photoproductions off
  nucleons in the quark model}.
\newblock {\em Phys. Rev. C}, 56:1099--1113, 1997.

\bibitem{Zhao:2002id}
Qiang Zhao, J.~S. Al-Khalili, Z.~P. Li, and R.~L. Workman.
\newblock {Pion photoproduction on the nucleon in the quark model}.
\newblock {\em Phys. Rev. C}, 65:065204, 2002.

\bibitem{LHCb:2018vuc}
Roel Aaij et~al.
\newblock {Observation of a new $\Xi_b^-$ resonance}.
\newblock {\em Phys. Rev. Lett.}, 121(7):072002, 2018.

\end{thebibliography}
\end{document}